\documentclass[%
 reprint,
 superscriptaddress,
 amsmath,amssymb,
 aps,
floatfix,
]{revtex4-2}

\usepackage{graphicx}
\usepackage{dcolumn}
\usepackage{bm}
\usepackage{hyperref}
\usepackage[mathlines]{lineno}
\usepackage{stmaryrd} 
\usepackage{float} 

\begin{document}

\preprint{APS/123-QED}

\title{Enforcing Analytic Constraints in Neural-Networks Emulating Physical Systems}

\author{Tom Beucler}
 \email{tom.beucler@gmail.com}
\affiliation{Department of Earth System Science, University of California, Irvine, CA, USA}%
\affiliation{Department of Earth and Environmental Engineering, Columbia University, New York, NY, USA}

\author{Michael Pritchard}
\affiliation{Department of Earth System Science, University of California, Irvine, CA, USA}%

\author{Stephan Rasp}
\affiliation{Technical University of Munich, Munich, Germany}

\author{Jordan Ott, Pierre Baldi}
\affiliation{Department of Computer Science, University of California, Irvine, CA, USA}

\author{Pierre Gentine}
\affiliation{Department of Earth and Environmental Engineering, Columbia University, New York, NY, USA}


\date{\today}

\begin{abstract}
Neural networks can emulate nonlinear physical systems with high accuracy, yet they may produce physically-inconsistent results when violating fundamental constraints. Here, we introduce a systematic way of enforcing nonlinear analytic constraints in neural networks via constraints in the architecture or the loss function. Applied to convective processes for climate modeling, architectural constraints enforce conservation laws to within machine precision without degrading performance. Enforcing constraints also reduces errors in the subsets of the outputs most impacted by the constraints. \\
\texttt{Main Repository: \url{https://github.com/raspstephan/CBRAIN-CAM}}\\ 
\texttt{Figures and Tables: \url{https://github.com/tbeucler/CBRAIN-CAM/blob/master/notebooks/tbeucler_devlog/042_Figures_PRL_Submission.ipynb}}
\end{abstract}

\maketitle


\section{Introduction}

Many fields of science and engineering (e.g., fluid dynamics, hydrology, solid mechanics, chemistry kinetics) have exact, often \textit{analytic}, closed-form constraints, i.e. constraints that can be explicitly written using analytic functions of the system's variables. Examples include translational or rotational invariance, conservation laws, or equations of state. While physically-consistent models should enforce constraints to within machine precision, data-driven algorithms often fail to satisfy well-known constraints that are not explicitly enforced. In particular, neural networks (NNs, \cite{baldi2021deep}), powerful regression tools for nonlinear systems, may severely violate constraints on individual samples while optimizing overall performance.  

Despite the need for physically-informed NNs for complex physical systems \citep{Reichstein2019,Bergen2019,Karpatne2017a,Willard2020}, enforcing \textit{hard} constraints \citep{Marquez-Neila2017} has been limited to physical systems governed by specific equations, such as advection equations \citep{Raissi2017,Bar-Sinai2019,DeBezenac2017}, Reynolds-averaged Navier-Stokes equations \citep{Ling2016,Wu2018}, boundary conditions of idealized flows \citep{Sun2020}, or quasi-geostrophic equations \citep{Bolton2019}. To address this gap, we introduce a systematic method to enforce analytic constraints arising in more general physical systems to within machine precision, namely the Architecture-Constrained NN or ACnet. We then compare ACnets to unconstrained (UCnets) and loss-constrained NNs (LCnets, in which soft constraints are added through a penalization term in the loss function \citep[e.g.,][]{Karpatne2017,Jia2019,Raissi2020}) in the particular case of climate modeling, where the system is high-dimensional and the constraints (such as mass and energy conservation) are few but crucial \citep{Beucler2019a}.

\section{Theory\label{sec:Theory}}

\subsection{Formulating the Constraints}

Consider a NN mapping an input vector $\boldsymbol{x}\in\mathbb{R}^{m}\ $ to an output vector $\boldsymbol{y}\in\mathbb{R}^{p}\ $. Enforcing constraints is easiest for linearly-constrained NNs, i.e. NNs for which the constraints $\left({\cal C}\right)\ $can be written as a linear system of rank $n$:
\begin{equation}
\left({\cal C}\right)\overset{\mathrm{def}}{=}\left\{ \boldsymbol{C}\left[\begin{array}{c}
\boldsymbol{x}\\
\boldsymbol{y}
\end{array}\right]=\boldsymbol{0}\right\} .\label{eq:Conservation}
\end{equation}
We call $\boldsymbol{C}\in\mathbb{R}^{n}\times\mathbb{R}^{m+p}\ $the constraints matrix, and use bold font for vectors and tensors to distinguish them from scalars. For the regression problem to have non-unique solutions, the number of independent constraints $ n$ has to be strictly less than $m+p $.

\begin{figure*}
\begin{centering}
\includegraphics{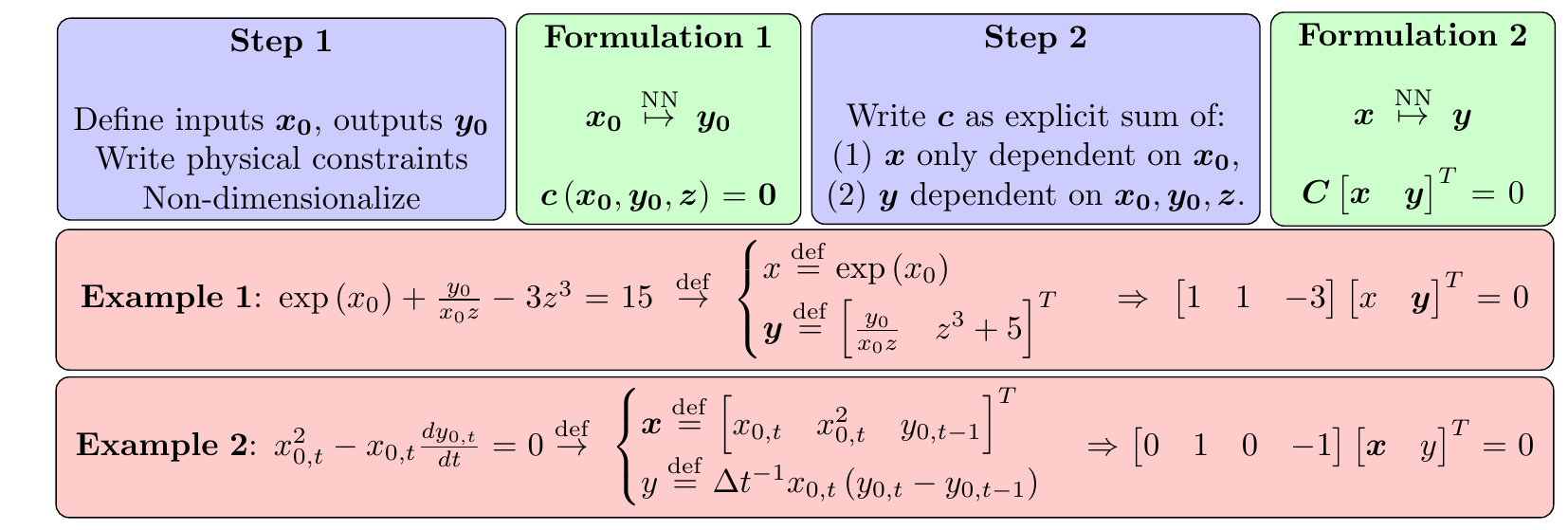}
\par\end{centering}

\caption{Framework to treat constrained regression problems using linearly-constrained NNs, with two examples: (1) A regression problem with one nonlinear constraint, and (2) a time-prediction problem with one differential nonlinear constraint that we discretize using a forward Euler method of timestep $\Delta t $. Note that the choice of $\boldsymbol{x},\boldsymbol{y} $, and $\boldsymbol{C} $ is not unique. \label{Fig01}}
\end{figure*}

In Figure \ref{Fig01}, we consider a generic regression problem subject to analytic constraints $\left({\cal C}\right) $ that may be nonlinear, and propose how to formulate a linearly-constrained NN. First, define the regression's inputs $ \boldsymbol{x_{0}}$ and outputs $ \boldsymbol{y_{0}}$, which respectively become the \textit{temporary} NN's features and targets. Then (\textbf{Formulation 1}), write the constraints $\left({\cal C}\right) $ as an identically zero function $ \boldsymbol{c}$ of the inputs, the outputs, and additional parameters $ \boldsymbol{z}$ the constraints may involve. We recommend non-dimensionalizing all variables to facilitate the design, interpretation, and performance of the loss function. While the function $ \boldsymbol{c}$ may be nonlinear, it can always be written as the sum of: (1) terms $ \boldsymbol{x}$ that \textit{only} depend on inputs and (2) terms $ \boldsymbol{y}$ that depend on inputs, outputs and additional parameters. Thus the constraints can be written as:  
\begin{equation}
\boldsymbol{c\left(x_{0},y_{0},z\right)}=\boldsymbol{C}\begin{bmatrix}\boldsymbol{x\left(x_{0}\right)}\\
\boldsymbol{y\left(x_{0},y_{0},z\right)}
\end{bmatrix},
\end{equation} 
where $\boldsymbol{C} $ is a matrix. Finally (\textbf{Formulation 2}), choose $\boldsymbol{x} $ and $\boldsymbol{y} $ as the NN's new inputs and outputs. If $\boldsymbol{x} $ and $\boldsymbol{y} $ are not bijective functions of $\boldsymbol{\left(x_{0},y_{0}\right)} $, add variables to the NN's inputs and outputs to recover $\boldsymbol{x_{0}} $ and $\boldsymbol{y_{0}} $ after optimization (e.g., we add $x_{0,t} $ and $y_{0,t-1} $ to $\boldsymbol{x} $ in \textbf{Example 2}). We are now in a position to build a computationally-efficient NN that satisfies the linear constraints $\left({\cal C}\right)$. 

\subsection{Enforcing the Constraints}

Consider a NN trained on preexisting
measurements of $\boldsymbol{x}\ $and $\boldsymbol{y} $. For simplicity's sake, we measure the quality of its output $\boldsymbol{y_{\mathrm{NN}}} $ using a standard mean-squared error (MSE) misfit:
\begin{equation}
\mathrm{MSE}\left(\boldsymbol{y_{\mathrm{Truth}}},\boldsymbol{y_{\mathrm{NN}}}\right)\overset{\mathrm{def}}{=}\left\Vert \boldsymbol{y_{\mathrm{Err}}}\right\Vert _{2}\overset{\mathrm{def}}{=}\frac{1}{p}\sum_{k=1}^{p}y_{\mathrm{Err},k}^{2},\label{eq:MSE}
\end{equation}
where we have introduced the error vector, defined as the difference between the NN's output and the ``truth'': 
\begin{equation}
\boldsymbol{y_{\mathrm{Err}}}\overset{\mathrm{def}}{=}\boldsymbol{y_{\mathrm{NN}}}-\boldsymbol{y_{\mathrm{Truth}}}.\label{Error_definition}
\end{equation}
In the reference case of an ``unconstrained network'' (UCnet), we optimize a multi-layer perceptron \citep[e.g., ][]{Jain1996,Gardner1998} using $\mathrm{MSE}\ $as its loss function $\cal{L} $. To enforce the constraints
$\left({\cal C}\right)\ $within NNs, we consider two options:

\textbf{(1) Constraining the loss function (LCnet, soft constraints)}: We first test a \textit{soft} penalization of the NN for violating physical
constraints using a penalty ${\cal P}$, defined as the mean-squared residual from
the constraints:
\begin{equation}
\begin{aligned}{\cal P}\left(\boldsymbol{x},\boldsymbol{y_{\mathrm{NN}}}\right) & \overset{\mathrm{def}}{=}\left\Vert \boldsymbol{C}\left[\begin{array}{c}
\boldsymbol{x}\\
\boldsymbol{y_{\mathrm{NN}}}
\end{array}\right]\right\Vert _{2},\\
 & =\frac{1}{n}\sum_{i=1}^{n}\left(\sum_{j=1}^{m}C_{ij}x_{j}+\sum_{k=1}^{p}C_{i\left(k+m\right)}y_{\mathrm{NN},k}\right)^{2},
\end{aligned}
\end{equation}
and given a weight $\alpha\in\left[0,1\right]\ $in the loss function ${\cal L}$:
\begin{equation}
{\cal L}\left(\alpha\right)=\alpha{\cal P}\left(\boldsymbol{x},\boldsymbol{y_{\mathrm{NN}}}\right)+\left(1-\alpha\right)\mathrm{MSE}\left(\boldsymbol{y_{\mathrm{Truth}}},\boldsymbol{y_{\mathrm{NN}}}\right).
\label{Loss function}
\end{equation}

\begin{figure}[b]
\begin{centering}
\includegraphics[width=0.5\textwidth]{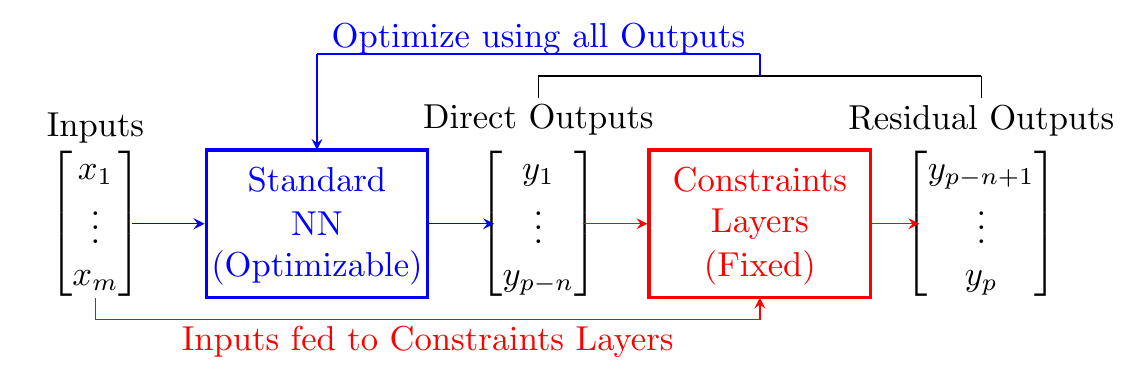}
\par\end{centering}

\caption{ACnet: Direct outputs are calculated using a standard NN, while the remaining outputs are calculated as residuals from the fixed constraints layers. \label{fig:NNA}}
\end{figure}

\textbf{(2) Constraining the architecture (ACnet, hard constraints):} Alternatively, we treat the constraints as \textit{hard} and augment a standard, optimizable NN with $n\ $fixed conservation layers that sequentially enforce the constraints $\left({\cal C}\right)\ $to within machine precision (Figure \ref{fig:NNA}), while keeping the $\mathrm{MSE}\ $as the loss function:
\begin{equation}
\left(\mathrm{ACnet}\right)\Rightarrow\begin{Bmatrix}\min\mathrm{MSE}\ \ \mathrm{s.t.}\ \ \boldsymbol{C}\begin{bmatrix}\boldsymbol{x} & \boldsymbol{y_{\mathrm{NN}}}\end{bmatrix}^{T}=\boldsymbol{0}\end{Bmatrix}
\end{equation}
The optimizable NN calculates a ``direct'' output whose size is $p-n$. We then calculate the remaining output's components of size $n$ as exact ``residuals'' from the constraints. Concatenating the ``direct'' and ``residual'' vectors results in the full output $\boldsymbol{y_{\mathrm{NN}}}\ $that satisfies the constraints to within machine precision. Since our loss uses the full output $\boldsymbol{y_{\mathrm{NN}}}$, the gradients of the loss function are passed through the constraints layers during optimization, meaning that the final NN's weights and biases depend on the constraints $\left({\cal C}\right)$. ACnet improves upon the common approach of calculating ``residual'' outputs \textit{after} training because ACnet exposes the NN to ``residual'' output data \textit{during} training (SM C.3). A possible implementation of the constraints layer uses custom (Tensorflow in our case) layers with fixed parameters that solve the system of equations $\left({\cal C}\right)\ $, in row-echelon form, from the bottom to the top row (SM B.1). Note that we are free to choose which outputs to calculate as ``residuals'', which introduces $n $ new hyperparameters (SM B.2).

\subsection{Linking Constraints to Performance}

Intuitively, we might expect the NNs' performance to improve once we enforce constraints arising in physical systems with few degrees of freedom, but this may not hold true with many degrees of freedom. We formalize the link between constraints and performance by: (1) decomposing the NN's prediction into the ``truth'' and error vectors following equation \ref{Error_definition}; and (2) assuming that constraints exactly hold for the ``truth'' (no errors in measurement). This yields:   
\begin{equation}
\boldsymbol{C}\left[\protect\begin{array}{c}
\boldsymbol{x}\protect\\
\boldsymbol{y_{\mathrm{NN}}}
\protect\end{array}\right]\protect\overset{\mathrm{def}}{=}\protect\overbrace{\boldsymbol{C}\left[\protect\begin{array}{c}
\boldsymbol{x}\protect\\
\boldsymbol{y_{\mathrm{Truth}}}
\protect\end{array}\right]}^{\boldsymbol{0}}+\boldsymbol{C}\left[\protect\begin{array}{c}
\boldsymbol{0}\protect\\
\boldsymbol{y_{\mathrm{Err}}}
\protect\end{array}\right].
\label{Decomposition_constraints_performances}
\end{equation}
Equation \ref{Decomposition_constraints_performances} relates how much the constraints are violated to the error vector. More explicitly, if we measure performance using the MSE, we may square each component of Equation \ref{Decomposition_constraints_performances}. The resulting equation links how much physical constraints are violated to the squared error for each constraint of index $i\in \llbracket1,n\rrbracket  $:

\begin{equation}
\begin{aligned}\underbrace{\left(\boldsymbol{C}\left[\begin{array}{c}
x\\
y_{\mathrm{NN}}
\end{array}\right]\right)_{i}^{2}}_{\mathrm{Physical\ constraints}} & =\underbrace{\sum_{k=1}^{p}C_{i\left(k+m\right)}^{2}y_{\mathrm{Err},k}^{2}}_{\mathrm{Squared-error}>0}\\
 & +\underbrace{\sum_{k=1}^{p}\sum_{l\neq k}C_{i\left(k+m\right)}C_{i\left(l+m\right)}y_{\mathrm{Err},k}y_{\mathrm{Err},l}}_{\mathrm{Cross-term}}
\end{aligned}
\label{Constraints_Performances}
\end{equation}

In ACnets, we strictly enforce physical constraints, setting the left-hand side of Equation \ref{Constraints_Performances} to 0, within numerical errors. As the squared error is positive-definite, the cross-term is always negative in ACnets as both terms sum up to 0. It is difficult to predict the cross-term before optimization, hence Equation \ref{Constraints_Performances} does not provide a-priori predictions of performance, even for ACnets. Instead, it links how much the NN violates constraints to how well it predicts outputs that appear in the constraints equations: the more negative the cross-term, the larger the squared error for a given violation of physical constraints.   

\section{Application\label{sec:Application}}

\subsection{Convective Parameterization for Climate Modeling}

The representation of subgrid-scale processes in coarse-scale, numerical models of the atmosphere, referred to as subgrid \textit{parameterization}, is a large source of error and uncertainty in numerical weather and climate prediction \citep[e.g.,][]{Palmer2005,Schneider2017}. Machine-learning algorithms trained on fine-scale, process-resolving models can improve subgrid parameterizations by faithfully emulating the effect of fine-scale processes on coarse-scale dynamics \citep[e.g.,][see Section 2 of \citet{Rasp2019a} for a detailed review]{Krasnopolsky2013,Gentine2018a,Rasp2018,Brenowitz2018}. The problem is that none of these parameterizations exactly follow conservation laws (e.g., conservation of mass, energy). This is critical for long-term climate projections, as the spurious energy production may both exceed the projected radiative forcing from greenhouse gases and result in large thermodynamic drifts or biases over a long time-period. Motivated by this shortcoming, we build a NN parameterization of convection and clouds that we \textit{constrain} to conserve 4 quantities: column-integrated energy, mass, longwave radiation, and shortwave radiation.

\subsection{Model and Data}

We use the Super-Parameterized Community Atmosphere Model 3.0 \citep{Khairoutdinov2005} to simulate the climate for two years in aquaplanet configuration \citep{Pritchard2014}, where the surface temperatures are fixed with a realistic equator-to-pole gradient \citep{Andersen2012}. Following \citep{Rasp2018}'s sensitivity tests, we use 42M samples from the simulation's first year to train the NN (training set) and 42M samples from the simulation's second year to validate the NN (validation set).
Since we use the validation set to adjust the NN's hyperparameters and avoid overfitting, we additionally introduce a test set using 42M different samples from the simulation's second year to provide an unbiased estimator of the NNs' performances. Note that each sample represents a single atmospheric column at a given time, longitude, and latitude.

\subsection{Formulating the Conservation Laws in a Neural Network}

The parameterization's goal is to predict the rate at which sub-grid convection vertically redistributes heat and water based on the current large-scale thermodynamic state. We group all variables describing the local climate in an input vector $\boldsymbol{x} $ of size 304 (5 vertical profiles with 30 levels each, prescribed large-scale conditions $\boldsymbol{\mathrm{LS}} $ for all profiles of size 150, and 4 scalars):
\begin{equation}
\boldsymbol{x}=\left[\begin{array}{ccc}
\left(\boldsymbol{q_{v}},\boldsymbol{q_{l}},\boldsymbol{q_{i}},\boldsymbol{T},\boldsymbol{v},\boldsymbol{\mathrm{LS}},p_{s},S_{0}\right) & \mathrm{SHF} & \mathrm{LHF}\end{array}\right]^{T},
\label{eq:x_def}
\end{equation}
where all variables are defined in SM A. We then concatenate the time-tendencies from convection and the additional variables involved in the conservation laws to form an output vector $\boldsymbol{y}\ $of size 216 (7 vertical profiles with 30 levels, followed by 6 scalars):
\begin{widetext}
\begin{equation}
\boldsymbol{y}=\left[\begin{array}{ccccccccccccc}
\boldsymbol{\dot{q}_{v}} & \boldsymbol{\dot{q}_{l}} & \boldsymbol{\dot{q}_{i}} & \boldsymbol{\dot{T}} & \boldsymbol{\dot{T}_{KE}} & \boldsymbol{\mathrm{lw}} & \boldsymbol{\mathrm{sw}} & \mathrm{LW_{t}} & \mathrm{LW_{s}} & \mathrm{SW_{t}} & \mathrm{SW_{s}} & P & P_{i}\end{array}\right]^{T},
\label{eq:y_def}
\end{equation}
\end{widetext}

We normalize all variables to the same units before non-dimensionalizing them using the constant $1\textnormal{W\ m}^{-2} $ (SM A.5). Finally, we derive the dimensionless conservation laws (SM A.1-A.4) and write them as a sparse matrix of size $4\times\left(304+218\right)\ $: 

\begin{widetext}
\begin{equation}
\boldsymbol{C}=\left[\begin{array}{cccccccccccccccc}
\boldsymbol{0} & 1 & \ell_{s} & -\ell_{s}\boldsymbol{\delta p} & -\ell_{f}\boldsymbol{\delta p} & \boldsymbol{0} & -\boldsymbol{\delta p} & \boldsymbol{\delta p} & \boldsymbol{0} & \boldsymbol{0} & -1 & 1 & 1 & -1 & -\ell_{f} & \ell_{f}\\
\boldsymbol{0} & 0 & 1 & -\boldsymbol{\delta p} & -\boldsymbol{\delta p} & -\boldsymbol{\delta p} & \boldsymbol{0} & \boldsymbol{0} & \boldsymbol{0} & \boldsymbol{0} & 0 & 0 & 0 & 0 & -1 & 0\\
\boldsymbol{0} & 0 & 0 & \boldsymbol{0} & \boldsymbol{0} & \boldsymbol{0} & \boldsymbol{0} & \boldsymbol{0} & \boldsymbol{\delta p} & \boldsymbol{0} & 1 & -1 & 0 & 0 & 0 & 0\\
\boldsymbol{0} & 0 & 0 & \boldsymbol{0} & \boldsymbol{0} & \boldsymbol{0} & \boldsymbol{0} & \boldsymbol{0} & \boldsymbol{0} & \boldsymbol{\delta p} & 0 & 0 & -1 & 1 & 0 & 0
\end{array}\right],
\label{eq:Conservation_matrix}
\end{equation}
\end{widetext}
that acts on $\boldsymbol{x}\ $and $\boldsymbol{y}\ $to yield Equation \ref{eq:Conservation}.

Each row of the constraints matrix $\boldsymbol{C}\ $describes a
different conservation law: The first row is column-integrated enthalpy conservation (here equivalent to energy conservation), the second row is column-integrated water conservation (here equivalent to mass conservation), the third row is column-integrated longwave radiation conservation and the last row is column-integrated shortwave radiation conservation. 

\subsection{Implementation}

We implement the three NN types and a multi-linear regression baseline using the Tensorflow library \citep{Abadi2016} version 1.13 with Keras \citep{chollet2015keras} version 2.2.4: (1) \textit{LCnets} for which we vary the weight $\alpha $ given to conservation laws from 0 to 1 (Equation \ref{Loss function}), (2) our reference \textit{ACnet}, and (3) \textit{UCnet}, i.e. an unconstrained LCnet of weight $\alpha=0 $. In our reference ACnet, we write the constraints layers in Tensorflow to solve the system of equations $\left({\cal C}\right) $ from bottom to top, and calculate surface tendencies as residuals of the conservation equations (SM B.1); switching the ``residual'' outputs to different vertical levels does not significantly change the validation loss nor the constraints penalty (SM B.3). After testing multiple architectures and activation functions (SM C.2), we chose 5 hidden layers of 512 nodes with leaky rectified linear-unit activations as our standard multi-layer perceptron architecture, resulting in $\sim $1.3M trainable parameters. We optimized the NN's weights and biases with the RMSprop optimizer \citep{tieleman2012lecture} for LCnets (because it was more stable than the Adam optimizer \citep{Kingma2014}), used Sherpa for hyperparameter optimizations \cite{hertel2020sherpa}, and saved the NN's state of minimal validation loss over 20 epochs.

\subsection{Results}

\begin{figure*}
\begin{centering}
\includegraphics[width=\textwidth]{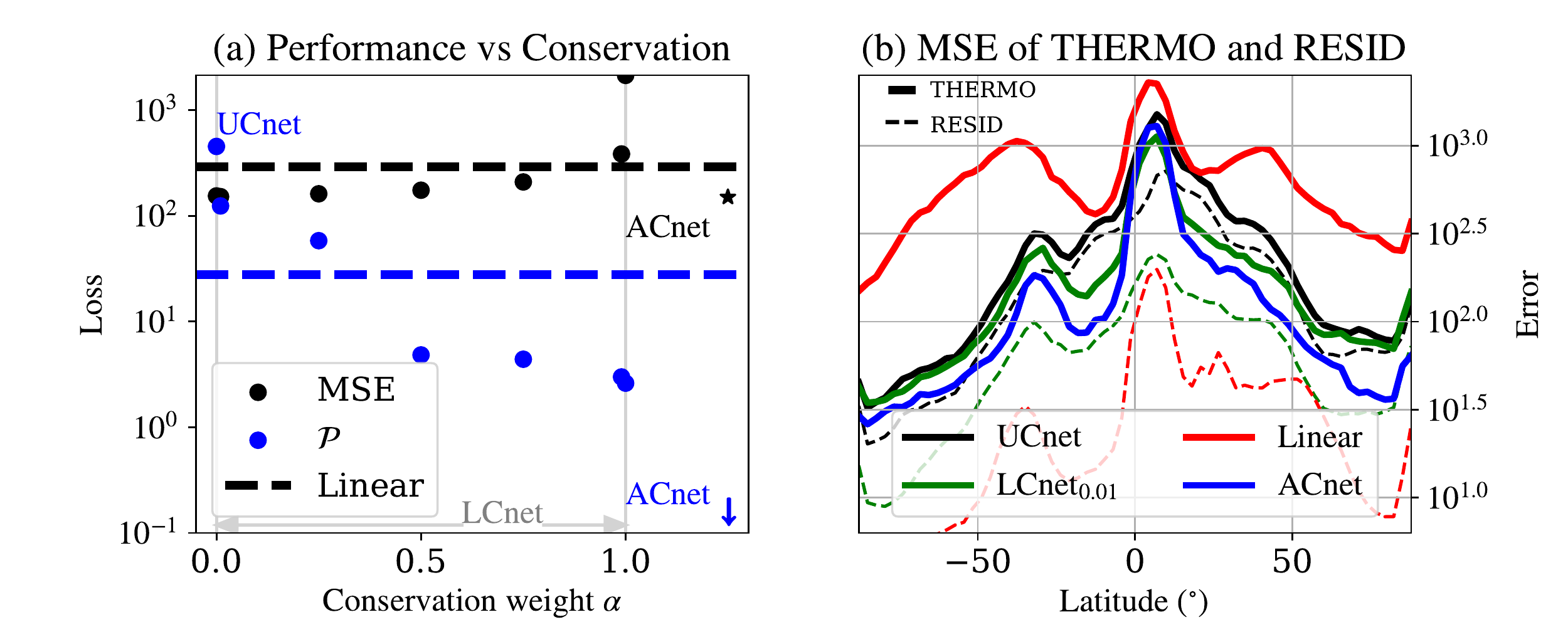}
\par\end{centering}

\caption{(a) MSE and $\cal{P} $ averaged over all samples of the test dataset for UCnet, LCnets of varying $\alpha $, and ACnet. The dashed lines indicate MSE and $\cal{P} $ for our multi-linear regression baseline. (b) Mean-squared error in the thermodynamic term (THERMO) and the enthalpy residual (RESID) versus latitude for our lowest-MSE NN in each category.\label{Fig3}}
\end{figure*}

In Figure \ref{Fig3}a, we compare mean performance (measured by MSE) and by how much physical constraints are violated (measured by $\cal{P} $) for the three NN types. As expected, we note a monotonic trade-off between performance and constraints as we increase $\alpha $ from 0 to 1 in the loss function. This trade-off is well-measured by MSE and $\cal P $ across the training, validation, and test sets (SM Table V). Interestingly, the physical constraints are easier to satisfy than reducing MSE in our case, likely because it is difficult to deterministically predict precipitation, which is strongly non-Gaussian, inherently stochastic, and whose error contributes to a large portion of MSE. Despite this, UCnet may violate physical constraints more than our multi-linear regression baseline. 

Our first key result is that \textit{ACnet performs nearly as well as our lowest-MSE UCnet on average} (to within 3$\% $) \textit{while satisfying constraints} to $\sim \left(10^{-9}\% \right) $ (SM C.1). This result holds across the training, validation and test sets (SM Table IV). In our case, ACnets perform slightly less well than UCnet because they are harder to optimize and the ``residual'' outputs exhibit systematically larger errors (SM B.2). This systematic, unphysical bias can be remedied by multiplying the weights of these ``residual'' outputs in the loss function (SM B.3) by a factor $\beta >1$ (SM Equation 12 and SM Figure 2). $\beta $ can be objectively chosen alongside the ``residual'' outputs via formal hyperparameter optimization (SM C.2).

In Figure \ref{Fig3}b, we compare how much the NNs violate column energy conservation (RESID) to the prediction of a variable that appears in that constraint: the total thermodynamic tendency in the enthalpy conservation equation (THERMO): 
\begin{equation}
\overbrace{\left(\boldsymbol{C}\left[\begin{array}{c}
x\\
y_{\mathrm{NN}}
\end{array}\right]\right)_{1}}^{\mathrm{RESID}} =\overbrace{\boldsymbol{\delta p}\cdot\boldsymbol{\left(\dot{T}_{\mathrm{KE}}-\dot{T}-\ell_{s}\dot{q}_{v}-\ell_{f}\dot{q}_{l}\right)}}^{\mathrm{THERMO}} + ...,
\end{equation}
where the ellipsis includes the surface fluxes, radiation, and precipitation terms. ACnet predicts THERMO more accurately than all NNs (full blue line) by an amount closely related to how much each NN violates enthalpy consevation (dashed lines), followed by LCnet (full green line). This yields our second key result: \textit{Enforcing constraints, whether in the architecture or the loss function, can systematically reduce the error of variables that appear in the constraints}. This result holds true across the training, validation, and test sets (SM Figure 4). However, possibly since our case has many degrees of freedom, it does not hold true for individual components of THERMO as their cross-term in Equation \ref{Constraints_Performances} is more negative for ACnet, nor does it hold for variables that are hard to predict deterministically (e.g., precipitation). Additionally, obeying conservation laws does not guarantee the ability to generalize well far outside of the training set, e.g. in the Tropics of a warmer climate (see Figure 3 of \citep{Beucler2020}). 
These results nuance the finding that physically constraining NNs systematically improves their generalization ability, which has been documented for machine learning emulation of low-dimensional idealized flows \citep{Sun2020,Willard2020}, and motivate physically-constraining machine-learning algorithms capable of stochastic predictions \citep{Wu2019} that are consistent across climates \citep{Beucler2020}. 


Finally, although the mapping presented in Section \ref{sec:Application} has linear constraints, ACnets can also be applied to nonlinearly constrained mappings by using the framework presented in Figure \ref{Fig01}. We give a concrete example in SM D, where we introduce the concept of ``conversion layers'' that transform nonlinearly constrained mappings into linearly-constrained mappings within NNs and without overly degrading performance (SM Table IX). Additionally, ACnets can be extended to incorporate inequality constraints on their ``direct'' outputs (by using positive-definite activation functions, discussed in SM E), making ACnets applicable to a broad range of constrained optimization problems. 


\begin{acknowledgments}
TB is supported by NSF grants OAC-1835769,  OAC-1835863, and AGS-1734164. PG acknowledges support from USMILE ERC synergy grant. The work of JO and PB is in part supported by grants NSF  1839429 and NSF NRT 1633631 to PB. We thank Eric Christiansen, Imme Ebert-Uphoff, Bart Van Merrienboer, Tristan Abbott, Ankitesh Gupta, and Derek Chang for advice. We also thank the meteorology department of LMU Munich and the Extreme Science and Engineering Discovery Environment supported by NSF grant number ACI-1548562 (charge numbers TG-ATM190002 and TG-ATM170029) for computational resources.
\end{acknowledgments}

\bibliography{main.bbl}

\end{document}


\preprint{APS/123-QED}

\title{Supplemental Material \\
Enforcing Analytic Constraints in Neural-Networks Emulating Physical Systems}

\author{Tom Beucler}
 \email{tom.beucler@gmail.com}
\affiliation{Department of Earth System Science, University of California, Irvine, CA, USA}%
\affiliation{Department of Earth and Environmental Engineering, Columbia University, New York, NY, USA}

\author{Michael Pritchard}
\affiliation{Department of Earth System Science, University of California, Irvine, CA, USA}%

\author{Stephan Rasp}
\affiliation{Technical University of Munich, Munich, Germany}

\author{Jordan Ott, Pierre Baldi}
\affiliation{Department of Computer Science, University of California, Irvine, CA, USA}

\author{Pierre Gentine}
\affiliation{Department of Earth and Environmental Engineering, Columbia University, New York, NY, USA}


\date{\today}


\maketitle

\setcounter{figure}{0}

The supplemental material (SM) has 5 Sections that can be read independently: (A) introduces the variables and constraints specific to our application, (B)  details the implementation of our architecture-constrained network, (C) compares different network types and architectures, (D) presents an example where \textit{nonlinear} constraints are enforced in neural networks emulating physical systems, and (E) discusses how to enforce \textit{inequality} constraints. 

\subsection*{A. Derivation of Dimensionless Conservation Equations}

The Super-Parameterized Community Atmosphere Model 3.0 embeds a convection-permitting model, namely the System for Atmospheric Modeling \citep{Khairoutdinov2003}, in each grid cell of the Community Atmosphere Model 3.0 \citep{Collins2006a}. In the absence of convective momentum transfer, our convection-permitting model conserves two quantities: liquid/ice water static energy and total water. We recast these conservation equations as an energy (A.1) and mass (A.2) conservation, before adding the conservation of longwave (A.3) and shortwave (A.4) radiation as our network predicts both radiative heating profiles and boundary fluxes at the top and bottom of the atmosphere, whose difference must match the mass-weighted vertical integral of the heating rate profiles. Finally, we non-dimensionalize all conservation equations in SM A.5. We define all variables in Table \ref{tab:Var_def}.

\begin{table}
\begin{tabular}{|c|c|}
\hline 
Variable & Name\tabularnewline
\hline 
\hline 
$\boldsymbol{\delta p}$ & Normalized differential pressure reference profile\tabularnewline
\hline 
$\ell_{f}$ & Normalized latent heat of fusion of water\tabularnewline
\hline 
$\ell_{s}$ & Normalized latent heat of sublimation of water\tabularnewline
\hline 
$\mathrm{LHF}$ & Latent heat flux\tabularnewline
\hline 
$\boldsymbol{\mathrm{LS}}$ & Large-scale forcings in water, temperature, velocity\tabularnewline
\hline 
$\boldsymbol{\mathrm{lw}}$ & Longwave heating rate profile\tabularnewline
\hline 
$\mathrm{LW_{s}}$ & Net surface longwave flux\tabularnewline
\hline 
$\mathrm{LW_{t}}$ & Net top-of-atmosphere longwave flux\tabularnewline
\hline 
$P$ & Total precipitation rate\tabularnewline
\hline 
$P_{i}$ & Solid precipitation rate\tabularnewline
\hline 
$p_{s}$ & Surface pressure\tabularnewline
\hline 
$S_{0}$ & Solar insolation\tabularnewline
\hline 
$\mathrm{SHF}$ & Sensible heat flux\tabularnewline
\hline 
$\boldsymbol{\mathrm{sw}}$ & Shortwave heating rate profile\tabularnewline
\hline 
$\mathrm{SW_{s}}$ & Net surface shortwave flux\tabularnewline
\hline 
$\mathrm{SW_{t}}$ & Net top-of-atmosphere shortwave flux\tabularnewline
\hline 
$\boldsymbol{T}$ & Absolute temperature profile\tabularnewline
\hline 
$\boldsymbol{\dot{T}}$ & Convective heating profile\tabularnewline
\hline 
$\boldsymbol{\dot{T}_{\mathrm{KE}}}$ & Heating from turbulent kinetic energy dissipation\tabularnewline
\hline 
$\boldsymbol{q_{i}}$ & Ice concentration profile\tabularnewline
\hline 
$\boldsymbol{\dot{q_{i}}}$ & Convective ice tendency profile\tabularnewline
\hline 
$\boldsymbol{q_{l}}$ & Liquid water concentration profile\tabularnewline
\hline 
$\boldsymbol{\dot{q_{l}}}$ & Convective liquid water tendency profile\tabularnewline
\hline 
$\boldsymbol{q_{v}}$ & Specific humidity profile\tabularnewline
\hline 
$\boldsymbol{\dot{q_{v}}}$ & Convective water vapor tendency profile\tabularnewline
\hline 
$\boldsymbol{v}$ & North-South velocity profile\tabularnewline
\hline 
$\boldsymbol{z}$ & Vertical level profile\tabularnewline
\hline
\end{tabular}
\caption{Definition of physical variables: Variables that depend on height are (boldfaced) vectors, referred to as ``profiles''.\label{tab:Var_def}}
\end{table}

\subsubsection*{A.1. Conservation of Energy}

We define the enthalpy $H\ $of an atmospheric column as the mass-weighted
vertical integral of the sum of its sensible heat and latent heat,
in $\mathrm{J\ m^{-2}}$, where ice is used as the reference
phase of zero energy:

\begin{equation}
H\overset{\mathrm{def}}{=}\int_{0}^{p_{s}}\frac{dp}{g}\underbrace{\left(c_{p}T+L_{s}q_{v}+L_{f}q_{l}\right)}_{h},\label{eq:MSE_definition}
\end{equation}

where $p\ $is atmospheric pressure in $\mathrm{Pa}$, $p_{s}\ $is
surface pressure in $\mathrm{Pa}$, $g\approx9.81\mathrm{m\ s^{-2}}\ $is
the gravity constant, $c_{p}\approx1.00.10^{3}\mathrm{J\ kg^{-1}\ K^{-1}}\ $is
the specific heat of water at constant pressure in standard atmospheric
conditions, $T\ $is the absolute temperature in $\mathrm{K}$,
$L_{s}\approx2.83.10^{6}\mathrm{J\ kg^{-1}}\ $is the latent heat
of sublimation of water in standard conditions, $q_{v}\ $is the specific
humidity or water vapor mass concentration in $\mathrm{kg/kg}$,
$L_{f}\approx3.34.10^{5}\mathrm{J\ kg^{-1}}\ $is the latent heat
of fusion of water in standard conditions, $q_{l}\ $is the liquid
water mass concentration in $\mathrm{kg/kg}$ and $h\ $is the
specific enthalpy in $\mathrm{J\ kg^{-1}}$. We isolate the
atmospheric column's time-tendency that is due to water phase changes
only $\left(\Delta_{\varphi}\right)\ $for each variable of Equation
\ref{eq:MSE_definition}:
\begin{subequations}
\begin{equation}
\Delta_{\varphi}h=-L_{f}\left(P-P_{i}\right),
\end{equation}

\begin{equation}
\begin{aligned}c_{p}\Delta_{\varphi}T & =\mathrm{-SHF}+\int_{0}^{p_{s}}\frac{dp}{g}c_{p}\left(\dot{T}+\dot{T}_{KE}\right)\\
 & \mathrm{+SW_{s}-SW_{t}+LW_{t}-LW_{s}},
\end{aligned}
\end{equation}

\begin{equation}
L_{v}\Delta_{\varphi}q_{v}=-\mathrm{LHF}+\int_{0}^{p_{s}}\frac{dp}{g}L_{v}\dot{q}_{v},
\end{equation}

\begin{equation}
\Delta_{\varphi}q_{l}=\int_{0}^{p_{s}}\frac{dp}{g}\dot{q}_{l},
\end{equation}
\end{subequations}
where $P\ $is the total surface precipitation rate in $\mathrm{kg\ m^{-2}\ s^{-1}}$,
$P_{i}\ $is the surface solid precipitation rate in $\mathrm{kg\ m^{-2}\ s^{-1}}$,
$\mathrm{SHF}\ $is the surface sensible heat flux in $\mathrm{W\ m^{-2}}$,
$\dot{T}\ $is the time-tendency of temperature in $\mathrm{K\ s^{-1}}$,
$\dot{T}_{KE}\ $is the time-tendency of temperature due to frictional
dissipation of kinetic energy in $\mathrm{K\ s^{-1}}$, $\mathrm{SW_{s}}\ $is
the net surface downwards shortwave radiative flux in $\mathrm{W\ m^{-2}}$,
$\mathrm{SW_{t}}\ $is the net top-of-atmosphere downwards shortwave radiative
flux in $\mathrm{W\ m^{-2}}$, $\mathrm{LW_{t}}\ $is the net
top-of-atmosphere upwards longwave radiative flux in $\mathrm{W\ m^{-2}}$,
$\mathrm{LW_{s}}\ $is the net surface upwards longwave radiative
flux in $\mathrm{W\ m^{-2}}$, $L_{v}\approx2.50.10^{6}\mathrm{J\ kg^{-1}}\ $is
the latent heat of vaporization of water in standard conditions, $\mathrm{LHF}\ $is
the surface latent heat flux in $\mathrm{W\ m^{-2}}$, $\dot{q}_{v}\ $is
the time tendency of specific humidity in $\mathrm{kg\ kg^{-1}\ s^{-1}}\ $and
$\dot{q}_{l}\ $is the time tendency of liquid water concentration
in $\mathrm{kg\ kg^{-1}\ s^{-1}}$. We then equate the time-tendencies
due to water phase changes in Equation \ref{eq:MSE_definition} to
form the equation describing the conservation of column enthalpy:

\begin{equation}
\Delta_{\varphi}h-c_{p}\Delta_{\varphi}T-L_{s}\Delta_{\varphi}q_{v}-L_{f}\Delta_{\varphi}q_{l}=0.\label{eq:Conservation_enthalpy}
\end{equation}

\subsubsection*{A.2. Conservation of Mass}

The conservation of water states that the change in total column water
concentration must balance the sources and sinks of water at the surface,
namely the surface evaporation and precipitation rates:

\begin{equation}
\int_{0}^{p_{s}}\frac{dp}{g}\left(\dot{q}_{v}+\dot{q}_{l}+\dot{q}_{i}\right)=\frac{\mathrm{LHF}}{L_{v}}-P.\label{eq:Conservation of water}
\end{equation}

\subsubsection*{A.3. Conservation of Longwave Radiation}

The conservation of longwave radiation states that the difference
between the top-of-atmosphere and surface longwave radiative fluxes
must balance the net longwave (or longwave) radiative cooling of
the atmospheric column to space:

\begin{equation}
\mathrm{LW_{t}-LW_{s}}=-\int_{0}^{p_{s}}\frac{dp}{g}c_{p}\mathrm{lw},\label{eq:Conservation of longwave}
\end{equation}

where $\mathrm{lw}\ $is the vertically-resolved temperature tendency due to longwave
heating in $\mathrm{K\ s^{-1}}$.

\subsubsection*{A.4. Conservation of Shortwave Radiation}

The conservation of shortwave radiation states that the difference between
the top-of-atmosphere insolation and the incoming shortwave radiation
at the surface must balance the net shortwave radiative
heating of the atmospheric column:

\begin{equation}
\mathrm{SW_{t}-SW_{s}}=\int_{0}^{p_{s}}\frac{dp}{g}c_{p}\mathrm{sw}.\label{eq:Conservation of shortwave}
\end{equation}

\subsubsection*{A.5. Non-dimensionalization of Conservation Equations}

We non-dimensionalize the conservation equations by converting all
tendencies to units $\mathrm{W\ m^{-2}}\ $before dividing them by
$1\mathrm{W\ m^{-2}}$. For numerical modeling purposes, each vertical
profile is discretized to 30 vertical levels $z$ of varying pressure
thickness $\delta{\cal P}_{z}$. This means that a continuous conservation
equations becomes a linear constraint on discrete variables; for instance
Equation \ref{eq:Conservation of shortwave} becomes:

\begin{equation}
\mathrm{SW_{t}-SW_{s}}=\sum_{z=1}^{30}\frac{\delta{\cal P}_{z}}{g}c_{p}\mathrm{sw}_{z}.\label{eq:Discretized_shortwave_conservation}
\end{equation}

To make Equation \ref{eq:Discretized_shortwave_conservation} non-dimensional,
we introduce a fixed, normalized differential pressure coordinate
$\delta p\ $to make the atmospheric pressure $p\ $non-dimensional:
$\forall z,\ \delta\widetilde{p}_{z}\overset{\mathrm{def}}{=}\delta{\cal P}_{z}/\delta p_{z}$.
This motivates the following non-dimensionalizations:

\begin{equation}
\mathrm{\widetilde{SW}_{t}\overset{def}{=}}\frac{\mathrm{SW_{t}}}{1\mathrm{W\ m^{-2}}}\ \ \mathrm{\widetilde{SW}_{s}\overset{def}{=}}\frac{\mathrm{SW_{s}}}{1\mathrm{W\ m^{-2}}}\ \ \widetilde{\mathrm{sw}}_{z}\overset{\mathrm{def}}{=}\frac{c_{p}\delta p_{z}\mathrm{sw}_{z}}{g},
\end{equation}

which leads to the simpler form of the shortwave conservation equation
presented in the main text:

\begin{equation}
\mathrm{\widetilde{SW}_{t}-\widetilde{SW}_{s}}=\sum_{z=1}^{30}\widetilde{\mathrm{sw}}_{z}\delta\widetilde{p}_{z}=\boldsymbol{\widetilde{\mathrm{sw}}}\cdot\boldsymbol{\delta\widetilde{p}}.
\end{equation}

For simplicity, the tildes are dropped in the main text and the following appendices.

\subsection*{B. Implementation of the Architecture-constrained Network}

In this section, we present our standard implementation of ACnets (B.1), which is application-specific, investigate the sensitivity of ACnets to the choice of ``residual'' outputs (B.2), and introduce a method to decrease the ``residual'' outputs' biases by preferentially weighting the loss function (B.3).

Note that there are multiple ways of implementing the ACnet defined in our manuscript. The first step involves choosing $\left(p-n\right) $ ``direct'' NN outputs calculated from the NN's weights and biases, and $n $ ``residual'' outputs calculated using the fixed constraints matrix $\boldsymbol{C} $. If a number $q\ $of outputs appear in the $n\ $constraints, this results in $\frac{n!}{q!\left(q-n\right)!} $ options to choose from. The second step involves calculating the $n $ residual outputs every time a batch is passed to the network. We can implement this by either:
\begin{itemize}
    \item adding custom layers to the NN that use the physical constraints $\left({\cal C}\right) $ to calculate the ``residual'' outputs, which are then concatenated with the ``direct'' outputs to pass an output vector of length $p $ to the loss function, or
    \item passing the ``direct'' outputs of length $\left(p-n\right) $ to the loss function and modifying the loss function's code to calculate the ``residual'' outputs within the loss function before e.g. calculating the mean-squared error.
\end{itemize}
We adopt the former approach below.

\subsubsection*{B.1. Standard Implementation}

In our standard implementation, we first output a vector of size $\left(218-4\right)=214$, before calculating the ``residual'' component of the output vector by solving the system of equations $\boldsymbol{C}\left[\begin{array}{cc}
\boldsymbol{x} & \boldsymbol{y}\end{array}\right]^{T}=\boldsymbol{0}\ $from bottom to top and within the network. All layers are implemented in Tensorflow using the functional application programming interface. Here, we choose to implement one layer per physical constraint, so that each layer solves one equation described by one row of the constraints matrix $\boldsymbol{C}$. We could equivalently have grouped all constraints in a single constraints layer, which would then have solved the entire system of equations described by the full constraints matrix $\boldsymbol{C}$. 

The first conservation layer $\left(\mathrm{CL}_{1}\right)\ $calculates the net shortwave
surface flux as a residual of the conservation of shortwave radiation
(last row of $\boldsymbol{C}$):
\begin{equation}
\sum_{i=1}^{30}\mathrm{sw_{z}}\delta p_{z}-\mathrm{SW_{t}+SW_{s}}=0.
\end{equation}

\begin{subequations}
\begin{equation}
\left(\mathrm{CL}_{1}\right)\ \ \underbrace{\mathrm{SW_{s}}}_{\mathrm{Residual}_{4}}=\mathrm{SW_{t}}-\sum_{z=1}^{30}\mathrm{sw}_{z}\delta p_{z}.
\end{equation}

The second conservation layer $\mathrm{\left(CL_{2}\right)}\ $calculates
the net longwave surface flux as a residual of the conservation of
longwave radiation (third row of $\boldsymbol{C}$):

\begin{equation}
\left(\mathrm{CL_{2}}\right)\ \ \underbrace{\mathrm{LW_{s}}}_{\mathrm{Residual}_{3}}=\mathrm{LW_{t}}-\sum_{z=1}^{30}\mathrm{lw}_{z}\delta p_{z}.
\end{equation}

The third conservation layer $\left(\mathrm{CL_{3}}\right)\ $calculates
the lowest-level specific humidity tendency as a residual of the conservation
of mass (second row of $\boldsymbol{C}$):

\begin{equation}
\begin{aligned} \left(\mathrm{CL}_{3}\right)\ \ \delta p_{30}\underbrace{\dot{q}_{v,30}}_{\mathrm{Residual}_{2}} & =\mathrm{LHF-P}-\sum_{z=1}^{29}\delta p_{z}\dot{q}_{v,z}\\ & -\sum_{z=1}^{30}\delta p_{z}\left(\dot{q}_{l,z}+\dot{q}_{i,z}\right).
\end{aligned}
\end{equation}

The fourth conservation layer $\mathrm{\left(CL_{4}\right)}\ $calculates
the lowest-level temperature tendency as a residual of the conservation
of energy (first row of $\boldsymbol{C}$):

\begin{equation}
\begin{aligned}\left(\mathrm{CL_{4}}\right)\ \ \delta p_{30}\overbrace{\dot{T}_{30}}^{\mathrm{Residual}_{1}} & =\mathrm{SHF}+\ell_{s}\mathrm{LHF}+\ell_{f}\left(P-P_{i}\right)\\
 & -\mathrm{LW_{t}+LW_{s}+SW_{t}-SW_{s}}\\
 & -\sum_{z=1}^{29}\delta p_{z}\dot{T}_{z}\\
 & +\sum_{z=1}^{30}\delta p_{z}\left(\dot{T}_{\mathrm{KE},z}-\ell_{s}\dot{q}_{v,z}-\ell_{f}\dot{q}_{l,z}\right).
\end{aligned}
\end{equation}
\end{subequations}

\subsubsection*{B.2. Sensitivity to Residual Index}

The indices of the output's components calculated as ``residuals'' -- i.e. which vertical level and specific variable is chosen to residually enforce the constraints on the column -- are new hyperparameters of ACnets: While we chose the lowest-level convective heating and moistening tendencies as the ``residuals'' of our reference ACnet, we are free to choose other vertical levels (e.g., top-of-atmosphere) or other variables (e.g., convective liquid or ice tendencies) as ``residuals''. To probe the sensitivity of ACnets to this unfamiliar hyperparameter choice, we train 5 ACnets with different vertical ``residual''-levels over 20 epochs, save their states of minimal validation loss to avoid overfitting, and report their performance and conservation properties in Table \ref{Tab:Residual_index}. Each ACnet is referred to as $q_{z1}T_{z2} $, where $z1 $ is the vertical level index (increasing downward from 0 at top-of-atmosphere to 29 at the surface) of the ``residual'' convective moistening for mass conservation, and $z2 $ the index of the ``residual'' convective heating for energy conservation. 

\begin{figure*}
\begin{centering}
\includegraphics[width=\textwidth]{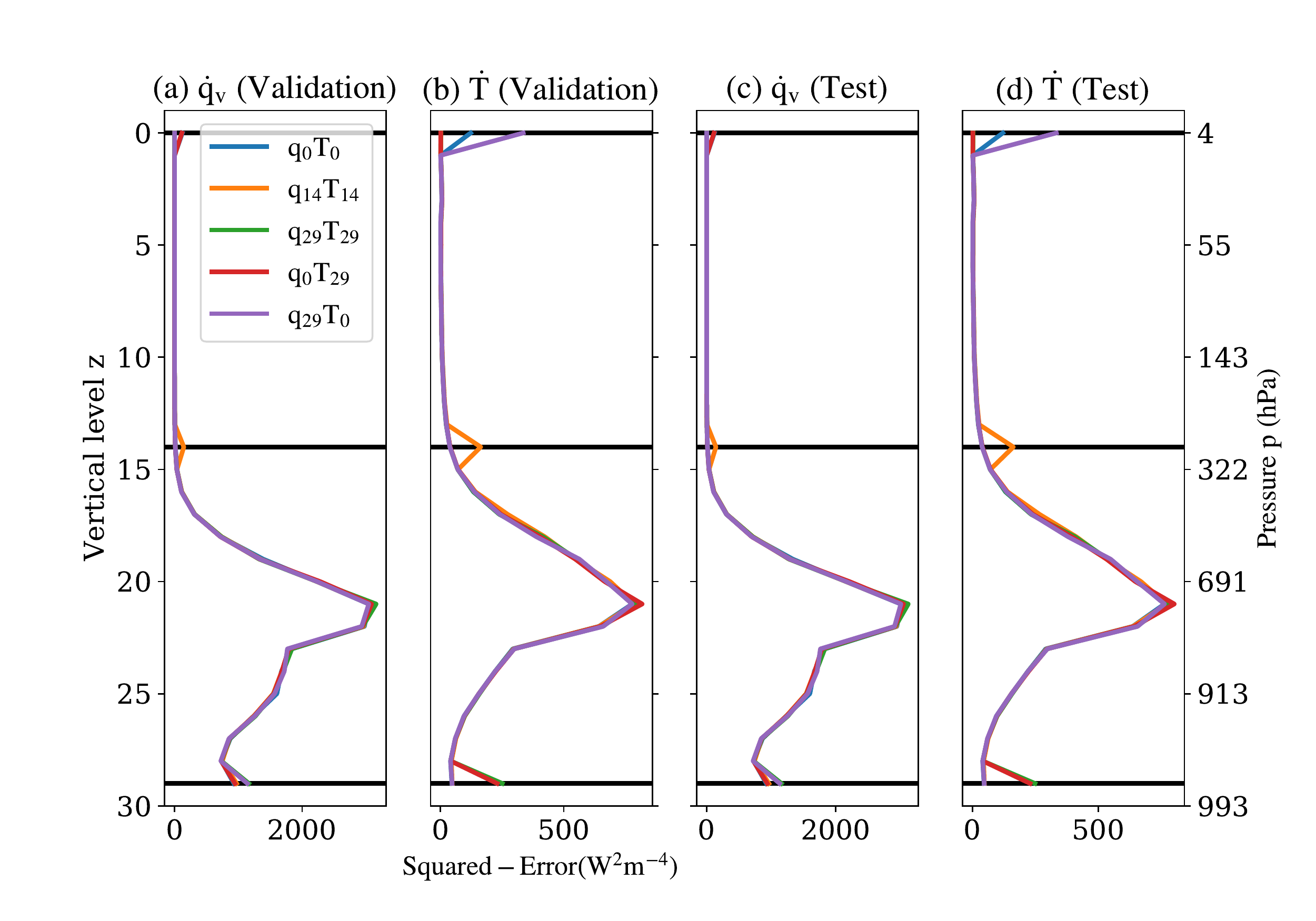}
\par\end{centering}
\caption{For various residual levels: Squared error in convective moistening $\boldsymbol{\dot{q}_{v}} $ and heating $\boldsymbol{\dot{T}} $ versus pressure, for the validation and test sets. We indicate ``residual'' levels where squared errors differ the most using black horizontal lines: The vertical level of index 0 is at the top of the atmosphere (0hPa), the vertical level of index 14 in the upper troposphere (274hPa), and the vertical level of index 29 at the surface (1000hPa). \label{Fig4}}
\end{figure*}

Reassuringly, all ACnets conserve column mass, energy and radiation to $\sim\left(10^{-9}\textnormal{W}^{2}\textnormal{m}^{-4}\right) $ over the training, validation and test sets (Table \ref{Tab:Residual_index}). However, a problem is apparent in the underlying vertical structure: Despite similar overall MSEs, ACnets' squared-error is larger at the ``residual'' vertical level. SM Figure \ref{Fig4} illustrates this by focusing on errors in the convective moistening and heating profiles averaged over the entire validation and test sets. Note that averaging the squared error over the entire validation set is equivalent to averaging the squared error in latitude, longitude, and over all days of our reference simulation used in the validation set. 

To first order, all ACnets exhibit similar vertical structures in their squared error: The squared error scales like the variance at each vertical level, except near the surface ($\mathrm{z}>22$) where a large portion of the signal cannot be determinisically predicted (see Section 4.2 of \cite{Gentine2018a} and the discussion on the performance for different horizontal grid spacings in \cite{Yuval2020a}). The scaling also does not hold in the upper atmosphere ($\mathrm{z}<10$) where the true convective moistening ${\dot{q}}_{v}$ variance is so small that the squared error in ${\dot{q}}_{v}$ is negligible and does not contribute to the overall MSE. Interestingly, each ACnet performs worse at its ``residual'' levels on both sets. For instance, $q_{14}T_{14} $ (orange line) always produces the largest error at its residual level (middle horizontal black line) relative to other ACnets. Similarly, $q_{29}T_{29} $ (green line) and $q_{0}T_{29} $ (red line) produce the largest convective heating error at the lowest model level (bottom horizontal black line). Although the sample-to-sample variability is large (e.g., MSE standard deviation in Table \ref{Tab:Residual_index}), latitude-pressure and longitude-pressure plots of convective heating and moistening errors show a systematic error increase at the ``residual level'' (not shown), confirming this is a robust error associated with the ``residual level'' hyperparameter. Additional errors in radiative fluxes and precipitation result in the total MSE, reported on the first, third, and fifth rows of Table \ref{Tab:Residual_index}. Since this unphysical vertical structure is a disadvantage of ACnets we propose a simple solution below.

\begin{table*}
\begin{centering}
{\small{}}%
\begin{tabular}{c|c|c|c|c|c|c}
{\small{}Dataset} & {\small{}Metric} & $q_{0}T_{0}$ & $q_{14}T_{14}$ & $q_{29}T_{29}$ & $q_{0}T_{29}$ & $q_{29}T_{0}$\tabularnewline
\hline
{\small{}Training} & {\small{}$\mathrm{MSE}$} & {\small{}$1.6\ 10^{+02}\pm9.6\ 10^{+02}$} & {\small{}$1.5\ 10^{+02}\pm9.7\ 10^{+02}$} & {\small{}$1.6\ 10^{+02}\pm9.7\ 10^{+02}$} & {\small{}$1.5\ 10^{+02}\pm9.6\ 10^{+02}$} & {\small{}$1.5\ 10^{+02}\pm9.5\ 10^{+02}$}\tabularnewline
set & {\small{}${\cal P}$} & {\small{}$7.8\ 10^{-10}\pm1.2\ 10^{-09}$} & {\small{}$7.6\ 10^{-10}\pm1.2\ 10^{-09}$} & {\small{}$8.1\ 10^{-10}\pm1.3\ 10^{-09}$} & {\small{}$8.0\ 10^{-10}\pm1.4\ 10^{-09}$} & {\small{}$7.6\ 10^{-10}\pm1.2\ 10^{-09}$}\tabularnewline
\cline{1-1}
{\small{}Validation} & {\small{}$\mathrm{MSE}$} & {\small{}$1.6\ 10^{+02}\pm9.8\ 10^{+02}$} & {\small{}$1.6\ 10^{+02}\pm9.9\ 10^{+02}$} & {\small{}$1.6\ 10^{+02}\pm9.9\ 10^{+02}$} & {\small{}$1.6\ 10^{+02}\pm9.8\ 10^{+02}$} & {\small{}$1.6\ 10^{+02}\pm9.8\ 10^{+02}$}\tabularnewline
set & {\small{}${\cal P}$} & {\small{}$7.8\ 10^{-10}\pm1.3\ 10^{-09}$} & {\small{}$7.7\ 10^{-10}\pm1.3\ 10^{-09}$} & {\small{}$8.2\ 10^{-10}\pm1.4\ 10^{-09}$} & {\small{}$8.0\ 10^{-10}\pm1.4\ 10^{-09}$} & {\small{}$7.6\ 10^{-10}\pm1.3\ 10^{-09}$}\tabularnewline
\cline{1-1}
{\small{}Test} & {\small{}$\mathrm{MSE}$} & {\small{}$1.6\ 10^{+02}\pm9.7\ 10^{+02}$} & {\small{}$1.6\ 10^{+02}\pm9.8\ 10^{+02}$} & {\small{}$1.6\ 10^{+02}\pm9.8\ 10^{+02}$} & {\small{}$1.5\ 10^{+02}\pm9.7\ 10^{+02}$} & {\small{}$1.6\ 10^{+02}\pm9.7\ 10^{+02}$}\tabularnewline
set & {\small{}${\cal P}$} & {\small{}$7.9\ 10^{-10}\pm1.3\ 10^{-09}$} & {\small{}$7.7\ 10^{-10}\pm1.3\ 10^{-09}$} & {\small{}$8.3\ 10^{-10}\pm1.4\ 10^{-09}$} & {\small{}$8.1\ 10^{-10}\pm1.5\ 10^{-09}$} & {\small{}$7.7\ 10^{-10}\pm1.3\ 10^{-09}$}\tabularnewline
\hline
\end{tabular}
\par\end{centering}{\small \par}

\caption{ACnets of varying residual levels for mass (m) and enthalpy (e) conservation, presented in SM Figure \ref{Fig4} (Mean MSE/Penalty $\pm $ Standard deviation)\label{Tab:Residual_index}}
\end{table*}

\subsubsection*{B.3 Decreasing the Residual Index Bias by Weighting the Loss Function}

A simple way to address the systematic bias induced at an ACnet's ``residual'' level is to increase the weight of these ``residual'' outputs in the loss function $\cal L $:
\begin{equation}
    {\cal L}\left(\beta\right)=\underbrace{\frac{1}{p-n}\sum_{k=1}^{p-n}y_{\mathrm{Err},k}^{2}}_{\mathrm{Direct\ Outputs\ MSE}}+\beta\times\underbrace{\frac{1}{n}\sum_{k=p-n+1}^{p}y_{\mathrm{Err},k}^{2}}_{\mathrm{Residual\ Outputs\ MSE}},
\end{equation}
where we have modified the MSE loss defined in Equation 3 by introducing a loss multiplier $\beta > 1 $.

\begin{figure*}
\begin{centering}
\includegraphics[width=\textwidth]{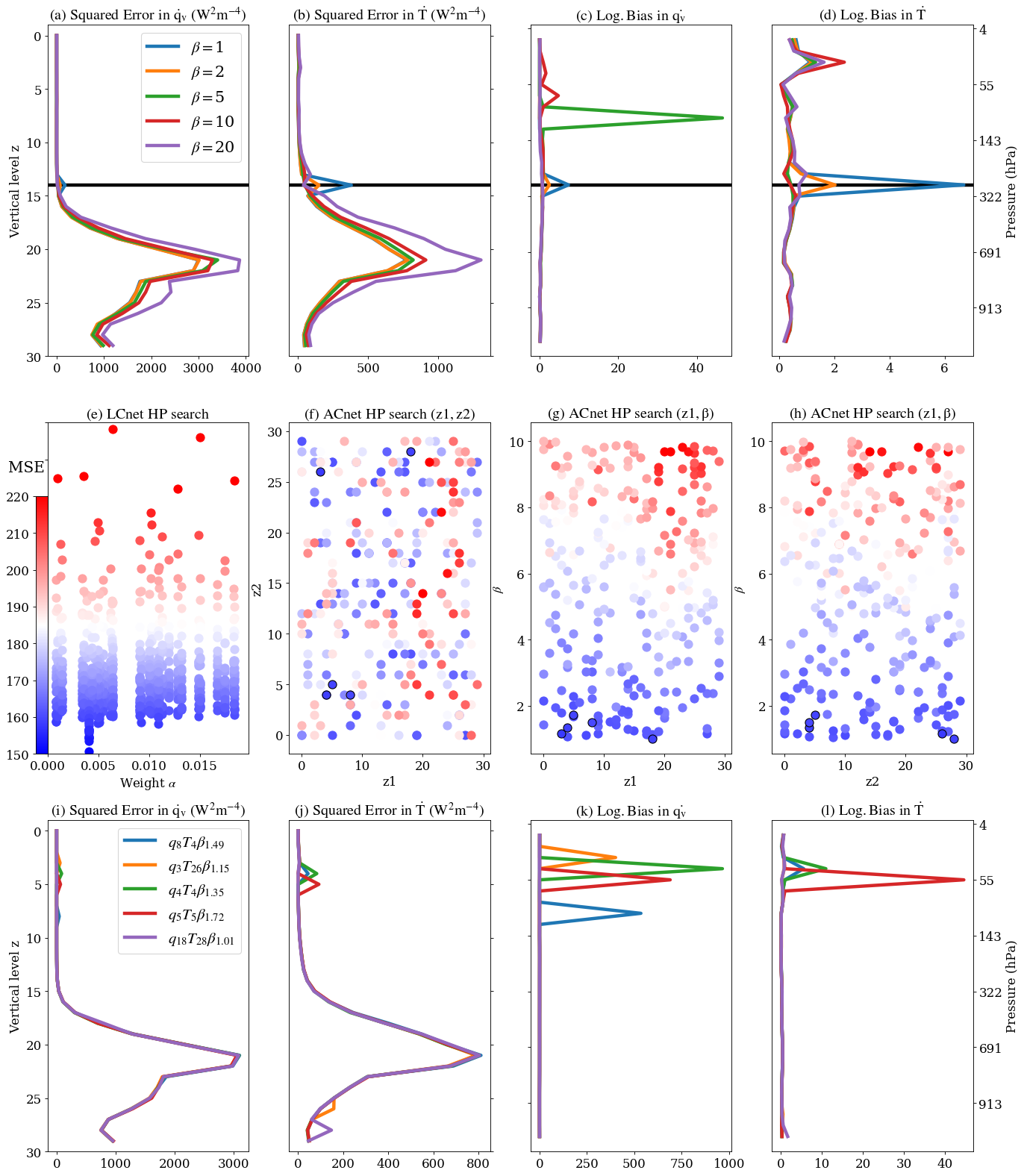}
\par\end{centering}
\caption{For various loss multipliers $\beta $: Squared error (a-b) and Log.Bias (c-d) in convective moistening $\boldsymbol{\dot{q}_{v}} $ and heating $\boldsymbol{\dot{T}} $ versus pressure, averaged over the test set. (e) LCnet HP search for low physical constraints weight $\alpha $, where dots are colored according to their MSE evaluated over the validation set. (f-h) ACnet HP search over $\left(\beta,\mathrm{z1},\mathrm{z2}\right)$, where the five NNs of minimal validation MSE are circled in black. For these five NNs of minimal validation MSE: Squared error (i-j) and Log.Bias (k-l) in convective moistening $\boldsymbol{\dot{q}_{v}} $ and heating $\boldsymbol{\dot{T}} $ versus pressure, averaged over the test set. \label{fig:SE_wloss}}
\end{figure*}

We implement the weighted loss as a custom Tensorflow loss and train ACnets of residual index $z=14 $ (i.e., $q_{14}T_{14} $) to test the sensitivity of the spurious error feature near 300 hPa (Fig \ref{Fig4}, orange line) with five different values of the loss multiplier $\beta $: 1, 2, 5, 10, and 20. We report the MSE and conservation penalty in Table \ref{tab:loss_multiplier}, and show the squared-error profiles averaged over the test set in SM Figures \ref{fig:SE_wloss}a and \ref{fig:SE_wloss}b. Reassuringly, all ACnets still conserve mass, energy and radiation to within machine precision. As $\beta $ increases, the spurious squared error anomaly at the ``residual'' index (horizontal black line at $z=14 $) becomes indistinguishable from the squared error at adjacent levels, successfully removing the anomaly. To quantitatively validate this observation and better visualize the ``residual'' bias, we introduce the absolute value of the logarithmic vertical gradient in the squared error $\boldsymbol{\mathrm{Log.bias}} $ as a proxy for the ``residual'' bias at each vertical level $\mathrm{z} $:
\begin{equation}
\begin{aligned}
\mathrm{Log.bias}_{z} & \overset{\mathrm{proxy}}{\approx}\frac{\left|\partial_{\mathrm{z}}y_{\mathrm{Err,z}}^{2}\right|}{y_{\mathrm{Err,z}}^{2}},\\    
 & \overset{\mathrm{numerics}}{\approx}\frac{\left|y_{\mathrm{Err,z+1}}^{2}-y_{\mathrm{Err,z}}^{2}\right|+\left|y_{\mathrm{Err,z}}^{2}-y_{\mathrm{Err,z-1}}^{2}\right|}{y_{\mathrm{Err,z+1}}^{2}+y_{\mathrm{Err,z-1}}^{2}},
\end{aligned}
\label{eq:log_bias}
\end{equation}
and depict it in SM Figures \ref{fig:SE_wloss}c and \ref{fig:SE_wloss}d. These figures underline the trade-off between the total MSE and the squared error at the residual level: As $\beta $ increases, the ``direct'' outputs are given less and less weight in the loss function $\cal L$, and while the bias at the ``residual'' levels decreases, the overall MSE increases. This is confirmed by (1) reading the MSE line of Table \ref{tab:loss_multiplier} from left to right, (2) the general increase of the squared-error profile with $\beta $ in SM Figures \ref{fig:SE_wloss}a and \ref{fig:SE_wloss}b; and (3) the large Log.bias introduced in the upper atmosphere ($z<10$) in SM Figures \ref{fig:SE_wloss}c and \ref{fig:SE_wloss}d.

In conclusion, introducing a moderate loss multiplier (e.g., $\beta=2 $) can systematically address the ``residual'' bias of ACnets at the cost of total MSE. $\beta $ can be seen as a new hyperparameter of ACnets, which can be tuned in conjunction with the ``residual'' indices to guarantee high performance and minimal ``residual'' biases as detailed in SM C.2.b.

\begin{table*}
\begin{centering}
{\small{}}%
\begin{tabular}{c|c|c|c|c|c|c}
{\small{}Validation} & {\small{}Metric} & $\beta=1$ & $\beta=2$ & $\beta=5$ & $\beta=10$ & $\beta=20$\tabularnewline
\hline
{\small{}Validation} & {\small{}$\mathrm{MSE}$} & {\small{}$1.6\ 10^{+02}\pm6.1\ 10^{+02}$} & {\small{}$1.6\ 10^{+02}\pm6.2\ 10^{+02}$} & {\small{}$1.7\ 10^{+02}\pm6.4\ 10^{+02}$} & {\small{}$1.9\ 10^{+02}\pm7.0\ 10^{+02}$} & {\small{}$2.4\ 10^{+02}\pm8.8\ 10^{+02}$}\tabularnewline
set & {\small{}${\cal P}$} & {\small{}$7.9\ 10^{-10}\pm8.1\ 10^{-09}$} & {\small{}$8.2\ 10^{-10}\pm8.6\ 10^{-09}$} & {\small{}$7.7\ 10^{-10}\pm7.7\ 10^{-09}$} & {\small{}$7.0\ 10^{-10}\pm7.4\ 10^{-09}$} & {\small{}$5.2\ 10^{-10}\pm5.4\ 10^{-09}$}\tabularnewline
\cline{1-1}
{\small{}Test} & {\small{}$\mathrm{MSE}$} & {\small{}$1.6\ 10^{+02}\pm6.0\ 10^{+02}$} & {\small{}$1.6\ 10^{+02}\pm6.0\ 10^{+02}$} & {\small{}$1.7\ 10^{+02}\pm6.2\ 10^{+02}$} & {\small{}$1.8\ 10^{+02}\pm6.9\ 10^{+02}$} & {\small{}$2.4\ 10^{+02}\pm8.6\ 10^{+02}$}\tabularnewline
set & {\small{}${\cal P}$} & {\small{}$7.9\ 10^{-10}\pm8.2\ 10^{-09}$} & {\small{}$8.2\ 10^{-10}\pm8.7\ 10^{-09}$} & {\small{}$7.8\ 10^{-10}\pm7.8\ 10^{-09}$} & {\small{}$7.1\ 10^{-10}\pm7.6\ 10^{-09}$} & {\small{}$5.3\ 10^{-10}\pm5.5\ 10^{-09}$}\tabularnewline
\cline{1-1}
\hline
\end{tabular}
\par\end{centering}{\small \par}

\caption{ACnets with varying loss multiplier $\beta $  given to the residual levels ($q_{14}T_{14}$) (Mean MSE/Penalty $\pm $ Standard deviation)\label{tab:loss_multiplier}}
\end{table*}

\subsection*{C. Comparison of Neural Network Types and Architectures}

In this section, we run sensitivity tests to probe the effect of our NN's characteristics on their performance and constraints penalty. Note that we distinguish NN types (i.e., LCnet, UCnet, ACnet; see SM C.1) from NN hyperparameters (i.e., number of layers, etc.; see SM C.2). In SM C.3, we compare ACnets that enforce constraints during training to NNs enforcing constraints after training.

\subsubsection*{C.1 Comparison between LCnets, UCnet and ACnet}

\begin{figure*}
\begin{centering}
\includegraphics[width=\textwidth]{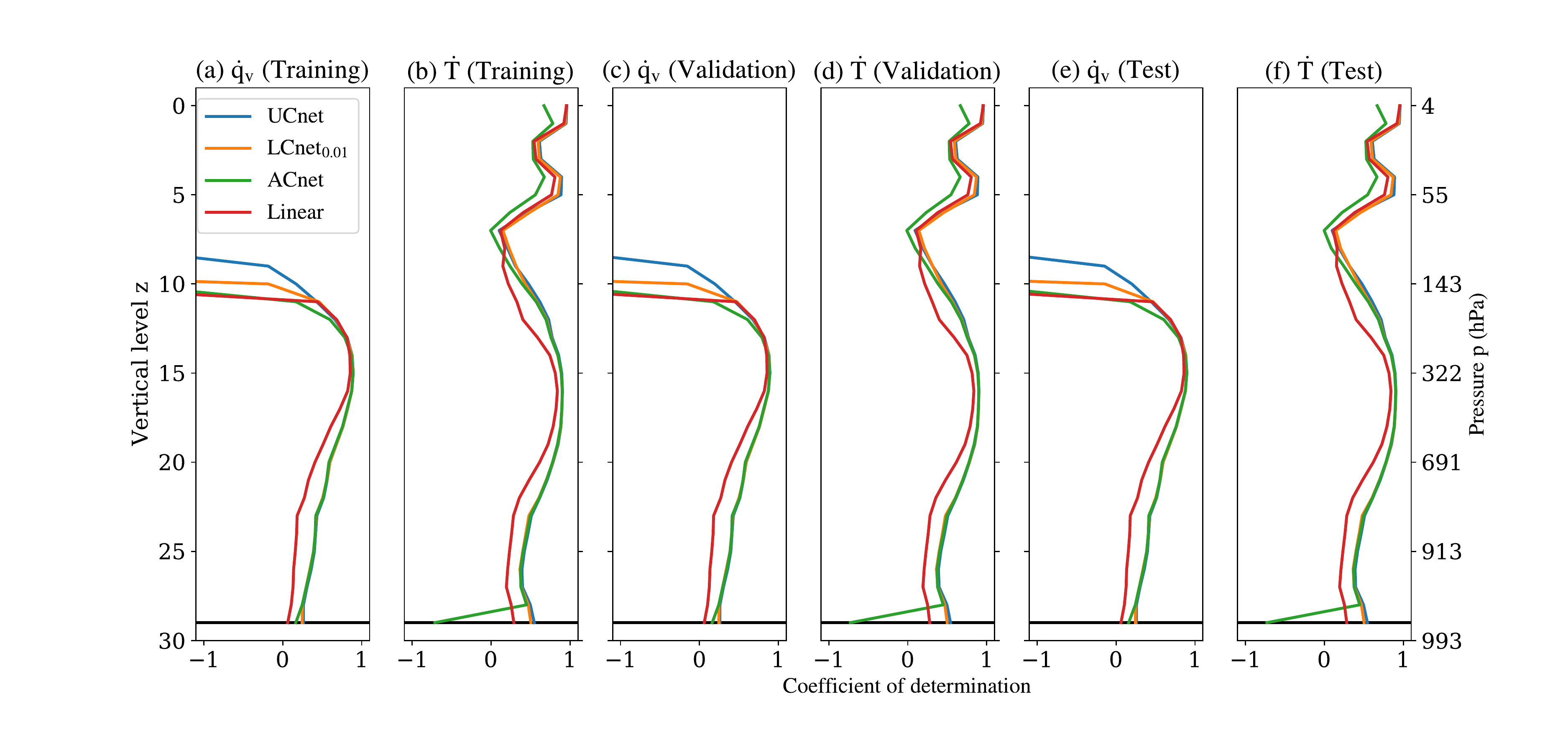}
\par\end{centering}
\caption{For various NN types and architectures: Coefficient of determination R$^{2} $ for convective moistening $\boldsymbol{\dot{q}_{v}} $ and heating $\boldsymbol{\dot{T}} $ versus pressure, for the training, validation and test sets. Note that squared errors in $\boldsymbol{\dot{q}_{v}} $ from levels 0 to 10 never exceed 10$^{-3} \%$ of the total squared error in $\boldsymbol{\dot{q}_{v}} $, making them irrelevant for our regression purposes. \label{fig:SE_Comparison}}
\end{figure*}

The performance and constraints penalties of the different NN types depicted in Figure 3b are compared in Table \ref{tab:Comparison}:
\begin{enumerate}
    \item ``Linear'' is the multi-linear regression baseline, derived by replacing all of UCnet's leaky rectified linear-unit activations with the identity function.
    \item ``UCnet'' is our best-performing NN, i.e. our NN of lowest MSE ($149\textnormal{ W}^{2}\textnormal{m}^{-4} $). Despite its high performance, it violates conservation laws more than our multi-linear regression, motivating ACnet and LCnet.
    \item ``LCnet ($\alpha=0.01) $'' is our LCnet with strictly-positive conservation weight $\alpha $ of lowest MSE ($151\textnormal{ W}^{2}\textnormal{m}^{-4} $). The $1\% $ conservation weight is enough to divide the mean penalty of ``UCnet'' by a factor 2.4 over the baseline validation dataset. Despite this improvement, samples at +1 standard deviation have conservation penalties of $\sim 50\textnormal{ W}^{2}\textnormal{m}^{-4} $, further motivating ACnet.
    \item ``ACnet'' is our reference ACnet described in SM B.1. Its MSE is $152\textnormal{ W}^{2}\textnormal{m}^{-4} $, which is only $3\textnormal{W}^{2}\textnormal{m}^{-4} $ more than our lowest-MSE UCnet.
\end{enumerate}
We present the performance and constraints penalties of LCnets of varying conservation weight in Table \ref{tab:LCnets} to better characterize the trade-off between performance and physical consistency depicted in Figure 3a. Note the large values of the conservation penalty's standard deviation for LCnets, illustrating the limits of enforcing constraints in the loss function. We further show the coefficient of determination R$^{2} $ for convective moistening and heating for the four different NN types in SM Figure \ref{fig:SE_Comparison}. NNs with nonlinear activation functions consistently outperform the multiple-linear regression baseline, while ACnets show a slight ``residual'' bias at the vertical level $z=29 $ (horizontal black line in SM Figure \ref{fig:SE_Comparison}).

\begin{figure*}
\begin{centering}
\includegraphics[width=1\textwidth]{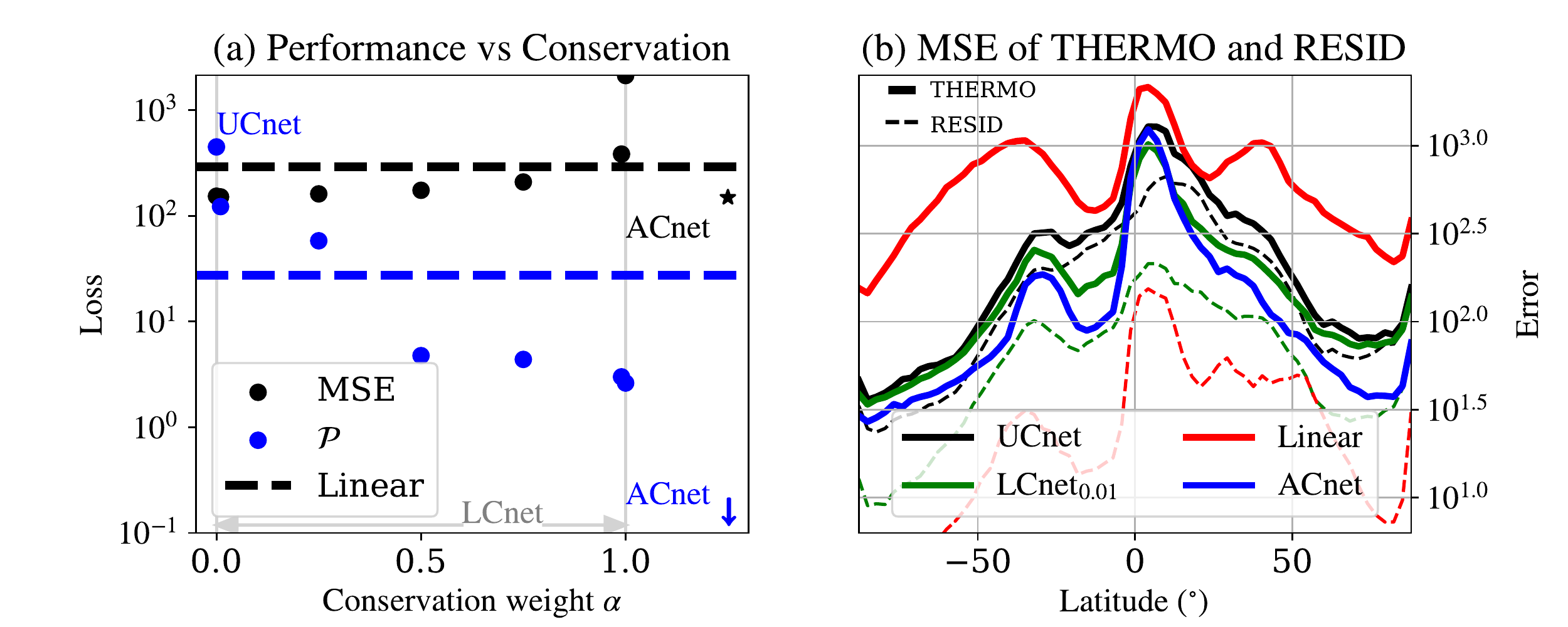}\\
\includegraphics[width=1\textwidth]{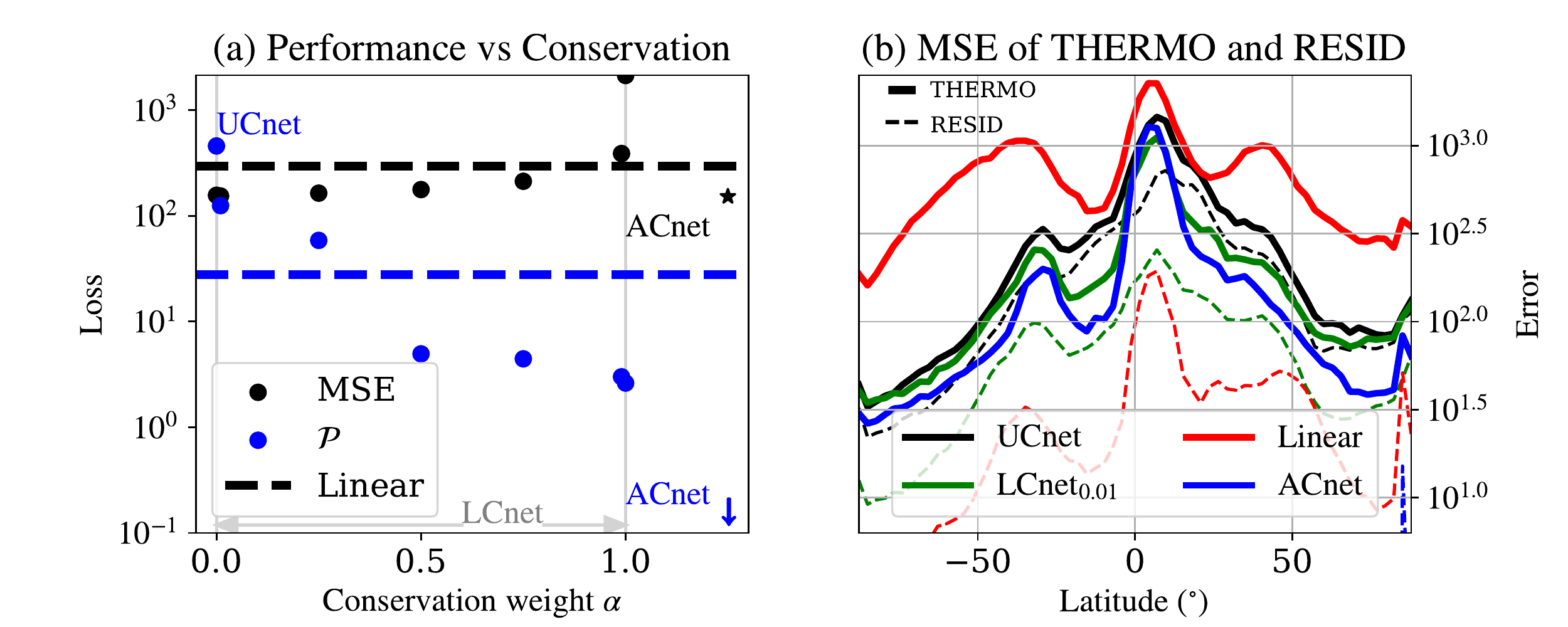}\\
\includegraphics[width=1\textwidth]{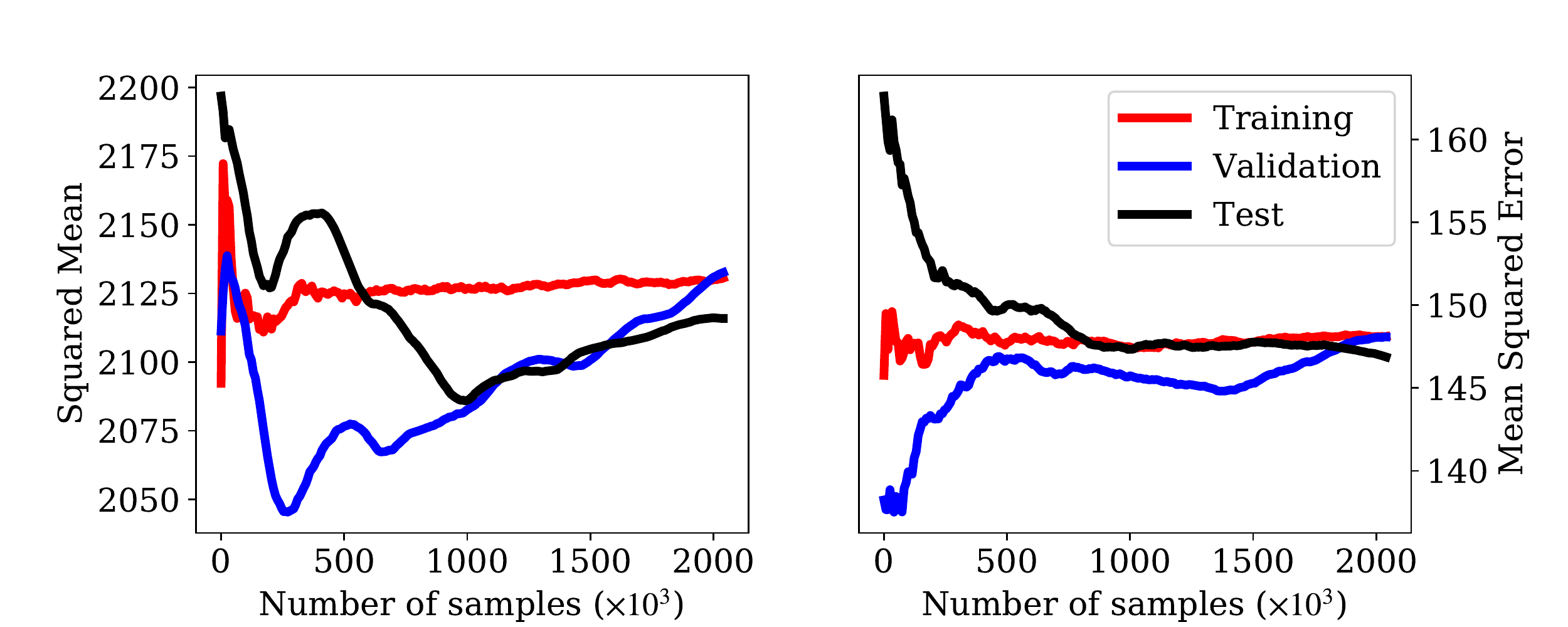}
\par\end{centering}
\caption{Figure 3 reproduced for the training (top row) and validation (middle row) sets. (Bottom row) Convergence of the squared mean output and UCnet's mean-squared error calculated over the training, validation, and test sets for large sample size. \label{fig:(Top-row)-Figure}}

\end{figure*}

Finally, as we produced Figure 3 using data from the test set, we reproduce Figure 3 using data from the training and validation sets below (SM Figure \ref{fig:(Top-row)-Figure}). The two top rows of SM Figure \ref{fig:(Top-row)-Figure} are nearly indistinguishable from Figure 3, confirming the robustness of our conclusions across the training, validation, and test sets. To explain why the two top rows are nearly indistinguishable in our case, we note that because all three sets are made of samples randomly drawn from a statistically-steady aquaplanet simulation, we expect statistical metrics to converge in the limit of large sample number. As this convergence is in contrast with standard training-validation splits in deep learning benchmarks (e.g., CIFAR-10 \citep{Krizhevsky2009}, ImageNet \citep{Russakovsky2015}), we illustrate it in the bottom row of SM Figure \ref{fig:(Top-row)-Figure} by plotting the squared mean of the true output vector (left) and the mean-squared error of UCnet (right) as a function of sample number from the training (red), validation (blue), and test (black) sets.

\begin{table*}
\begin{centering}
{\small{}}%
\begin{tabular}{c|c|c|c|c|c|c}
{\small{}Validation} & {\small{}Metric} & {\small{}Linear} & $\mathrm{UC_{net}}$ & $\mathrm{LC_{net}\left(\alpha=0.01\right)}$ & $\mathrm{AC_{net}}$\tabularnewline
\hline
{\small{}Training} & {\small{}$\mathrm{MSE}$} & {\small{}$2.9\ 10^{+02}\pm1.6\ 10^{+03}$} & {\small{}$1.5\ 10^{+02}\pm9.2\ 10^{+02}$} & {\small{}$1.5\ 10^{+02}\pm9.2\ 10^{+02}$} & {\small{}$1.5\ 10^{+02}\pm9.4\ 10^{+02}$}\tabularnewline
set & {\small{}${\cal P}$} & {\small{}$2.7\ 10^{+01}\pm2.2\ 10^{+01}$} & {\small{}$8.9\ 10^{+01}\pm8.0\ 10^{+01}$} & {\small{}$3.7\ 10^{+01}\pm2.8\ 10^{+01}$} & {\small{}$8.3\ 10^{-10}\pm1.4\ 10^{-09}$}\tabularnewline
\cline{1-1}
{\small{}Validation} & {\small{}$\mathrm{MSE}$} & {\small{}$3.0\ 10^{+02}\pm1.7\ 10^{+03}$} & {\small{}$1.5\ 10^{+02}\pm9.4\ 10^{+02}$} & {\small{}$1.5\ 10^{+02}\pm9.5\ 10^{+02}$} & {\small{}$1.5\ 10^{+02}\pm9.6\ 10^{+02}$}\tabularnewline
set & {\small{}${\cal P}$} & {\small{}$2.8\ 10^{+01}\pm2.3\ 10^{+01}$} & {\small{}$9.1\ 10^{+01}\pm8.2\ 10^{+01}$} & {\small{}$3.8\ 10^{+01}\pm2.8\ 10^{+01}$} & {\small{}$8.4\ 10^{-10}\pm1.5\ 10^{-09}$}\tabularnewline
\cline{1-1}
{\small{}Test} & {\small{}$\mathrm{MSE}$} & {\small{}$2.9\ 10^{+02}\pm1.7\ 10^{+03}$} & {\small{}$1.5\ 10^{+02}\pm9.4\ 10^{+02}$} & {\small{}$1.5\ 10^{+02}\pm9.4\ 10^{+02}$} & {\small{}$1.5\ 10^{+02}\pm9.6\ 10^{+02}$}\tabularnewline
set & {\small{}${\cal P}$} & {\small{}$2.8\ 10^{+01}\pm2.4\ 10^{+01}$} & {\small{}$9.0\ 10^{+01}\pm8.2\ 10^{+01}$} & {\small{}$3.8\ 10^{+01}\pm2.8\ 10^{+01}$} & {\small{}$8.5\ 10^{-10}\pm1.5\ 10^{-09}$}\tabularnewline
\hline
\end{tabular}
\par\end{centering}{\small \par}

\caption{NNs presented in Figure 3b (Mean MSE/Penalty $\pm $ Standard deviation)\label{tab:Comparison}}
\end{table*}

\begin{table*}
\begin{centering}
{\small{}}%
\begin{tabular}{c|c|c|c|c|c|c}
{\small{}Validation} & {\small{}Metric} & $\alpha=0$ & $\alpha=0.25$ & $\alpha=0.5$ & $\alpha=0.75$ & $\alpha=0.99$\tabularnewline
\hline
{\small{}Training} & {\small{}$\mathrm{MSE}$} & {\small{}$1.5\ 10^{+02}\pm9.6\ 10^{+02}$} & {\small{}$1.6\ 10^{+02}\pm9.9\ 10^{+02}$} & {\small{}$1.7\ 10^{+02}\pm1.0\ 10^{+03}$} & {\small{}$2.1\ 10^{+02}\pm1.2\ 10^{+03}$} & {\small{}$3.8\ 10^{+02}\pm1.8\ 10^{+03}$}\tabularnewline
set & {\small{}${\cal P}$} & {\small{}$4.5\ 10^{+02}\pm4.5\ 10^{+02}$} & {\small{}$5.8\ 10^{+01}\pm6.2\ 10^{+01}$} & {\small{}$4.8\ 10^{+00}\pm4.7\ 10^{+00}$} & {\small{}$4.4\ 10^{+00}\pm2.6\ 10^{+00}$} & {\small{}$3.0\ 10^{+00}\pm1.9\ 10^{+00}$}\tabularnewline
\cline{1-1}
{\small{}Validation} & {\small{}$\mathrm{MSE}$} & {\small{}$1.6\ 10^{+02}\pm9.8\ 10^{+02}$} & {\small{}$1.6\ 10^{+02}\pm1.0\ 10^{+03}$} & {\small{}$1.8\ 10^{+02}\pm1.1\ 10^{+03}$} & {\small{}$2.1\ 10^{+02}\pm1.2\ 10^{+03}$} & {\small{}$3.9\ 10^{+02}\pm1.8\ 10^{+03}$}\tabularnewline
set & {\small{}${\cal P}$} & {\small{}$4.6\ 10^{+02}\pm4.7\ 10^{+02}$} & {\small{}$5.9\ 10^{+01}\pm6.4\ 10^{+01}$} & {\small{}$4.9\ 10^{+00}\pm4.9\ 10^{+00}$} & {\small{}$4.4\ 10^{+00}\pm2.7\ 10^{+00}$} & {\small{}$3.0\ 10^{+00}\pm1.9\ 10^{+00}$}\tabularnewline
\cline{1-1}
{\small{}Test} & {\small{}$\mathrm{MSE}$} & {\small{}$1.5\ 10^{+02}\pm9.7\ 10^{+02}$} & {\small{}$1.6\ 10^{+02}\pm1.0\ 10^{+03}$} & {\small{}$1.7\ 10^{+02}\pm1.0\ 10^{+03}$} & {\small{}$2.1\ 10^{+02}\pm1.2\ 10^{+03}$} & {\small{}$3.8\ 10^{+02}\pm1.8\ 10^{+03}$}\tabularnewline
set & {\small{}${\cal P}$} & {\small{}$4.5\ 10^{+02}\pm4.7\ 10^{+02}$} & {\small{}$5.8\ 10^{+01}\pm6.5\ 10^{+01}$} & {\small{}$4.8\ 10^{+00}\pm4.8\ 10^{+00}$} & {\small{}$4.4\ 10^{+00}\pm2.6\ 10^{+00}$} & {\small{}$3.0\ 10^{+00}\pm1.9\ 10^{+00}$}\tabularnewline
\hline
\end{tabular}
\par\end{centering}{\small \par}

\caption{LCnets of varying weight $\alpha $, presented in Figure 3a (Mean MSE/Penalty $\pm $ Standard deviation)\label{tab:LCnets}}
\end{table*}

\subsubsection*{C.2 Formal Hyperparameter Optimization}

To optimize the performance of our NN emulators and address some of their recurring biases, we conduct four distinct hyperparameter (HP) searches. We formally implement each HP search using SHERPA \citep{Hertel2018}, a Python library for HP tuning. We choose to conduct \textit{random} searches to explore a wide range of HP settings and avoid making assumption about the structure of the HP search problem \citep{Bergstra2012}. Each search involves training more than 200 trial models for at least 20 epochs to find the NN yielding the lowest MSE over the validation set. We follow the structure of the main text and first conduct one UCnet HP search (C.2.a), followed by two LCnet (C.2.b) and one ACnet (C.2.c) searches.



\paragraph{UCnet HP Optimization}
\   
     
\begin{figure*}
\begin{centering}
\includegraphics[width=\textwidth]{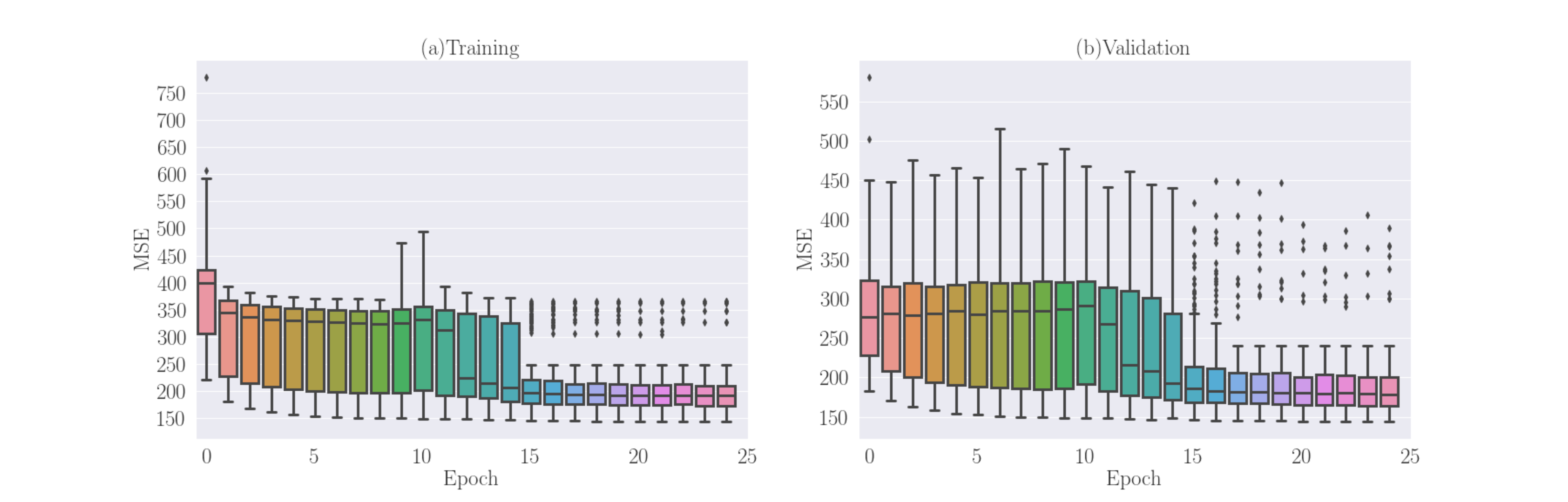}
\par\end{centering}
\caption{Training and validation MSE versus number of epochs from more than 200 models. }
\label{fig:mse_epochs}
\end{figure*}

We follow common practices to tune UCnets and conduct our searches over the number of layers, the number of nodes per layer, the properties of the activation function, the presence of batch normalization, and the dropout coefficient (see Table \ref{hyperparameter-table} for the range of available options during the search). The parameters of the best-performing UCnet are listed in Table \ref{tab:bestconfig_alpha0}, while the results notebook can be found at: \url{https://github.com/jordanott/CBRAIN-CAM/blob/master/notebooks/tbeucler_devlog/hp_opt_conservation/NormalMSE.ipynb}. As tuning LCnets and ACnets involves additional HPs ($\alpha$, $\beta $, and the residual indices), we use Table \ref{tab:bestconfig_alpha0} to guide our choice of baseline HPs, which we round to: 5 layers of 512 nodes, Leaky ReLU coefficient of 0.3, no dropout and no batch normalization. Finally, we use this first HP search to objectively choose an appropriate number of epochs. SM Figure \ref{fig:mse_epochs} shows training and validation MSE curves for more than 200 models. Both metrics plateau after 20 epochs, indicating this is an appropriate time to terminate training.

\begin{table}
\footnotesize
\centering
\caption{Hyperparameter Space}
\label{hyperparameter-table}
\begin{tabular}{@{}ll@{}ll@{}}
\toprule
Name                        & Options    & Parameter Type         \\ 
Batch Normalization        & {[}yes, no{]}        &   Choice         \\
Dropout                     & {[}0, 0.25{]}        &   Continuous  \\
Leaky ReLU coefficient      & {[}0 - 0.4{]}        &   Continuous      \\
Learning Rate               & {[}0.0001 - 0.01{]}  &   Continuous (log) \\
Nodes per Layer             & {[}300 - 700{]}      &   Discrete   \\
Number of layers            & {[}3 - 8{]}         &   Discrete       \\

\end{tabular}
\end{table}

\begin{table}
\footnotesize
    \centering
    \caption{Best HP configuration for UCnets $\left(\alpha=0\right) $}
    \begin{tabular}{ll}
    \toprule
    Batch Normalization         &   No                          \\
    Dropout                     &   0.00975                      \\
    Leaky ReLU coefficient      &   0.25373                      \\
    Learning Rate               &   0.000977                    \\
    Number of layers            &   5                          \\
    Nodes per Layer             &  {[}625, 517, 543, 538, 692{]}\\
    \end{tabular}
\label{tab:bestconfig_alpha0}
\end{table}

Using the baseline HPs identified via the UCnet HP search, we are now interested in the values of $\alpha$ for LCnets and $\left(\beta,\mathrm{z1},\mathrm{z2}\right)$ for ACnets that yield the best performance. In the following paragraphs, we quantify performance using MSE averaged over the validation set for consistency, but note that HP searches are flexible enough to target any performance metric that can be calculated from the NN's input and output vectors.

\paragraph{LCnet HP Optimization}
\   
    
Our first LCnet HP search (not shown) indicates that as suggested in Figure 3, SM Figure \ref{fig:(Top-row)-Figure}, and Table \ref{tab:LCnets}, MSE typically increases with $\alpha $ for LCnets. This means that low values of $\alpha $ systematically lead to the lowest MSEs if LCnets are trained for enough epochs: While our LCnets of lowest MSEs use $\alpha \approx 0.01$, their MSE is still larger than our best UCnet's MSE. This suggests a dominant trade-off between performance and physical constraints. For completeness, we conduct a second HP search restricted to $\alpha \in ]0,0.02[$ that we depict in Figure \ref{fig:SE_wloss}e and identify $\alpha = 4.1\ 10^{-4}$ as the weight leading to the minimal validation MSE of $151$ W$^{2}$m$^{-4}$. 

\paragraph{ACnet HP Optimization}
\   
    
Finally, we conduct a HP search over $\left(\beta,\mathrm{z1},\mathrm{z2}\right)\in]1,10]\times \llbracket0,29\rrbracket^{2}$ to find the set of loss multiplier and residual indices that maximize performance while still addressing biases at the ``residual'' vertical level. While good performance can be obtained for all values of $\mathrm{z_{2}}$, it is preferable to use $\mathrm{z_{1}}\apprle20$ (Figures \ref{fig:SE_wloss}e). Consistently with Figures \ref{fig:SE_wloss}a-d, there is a general trade-off between overall MSE and decreasing the ``residual'' bias, which suggests using $\beta\apprle2$ to maintain competitive performance (Figures \ref{fig:SE_wloss}f-g). For completeness, we identify the 5 NNs of lowest validation MSE (circled in black) and evaluate their squared error and Log.Bias over the test set (Figures \ref{fig:SE_wloss}i-l). 

All NNs exhibit biases at their ``residual'' levels, which we list in Figure \ref{fig:SE_wloss}i's legend. We can conveniently visualize these ``residual'' biases using the Log.Bias defined in equation \ref{eq:log_bias}, except near the surface in which case we may directly use the squared error. Finally, we note that the highest-performing NNs use larger $\beta $ for lower $\mathrm{z2}$, confirming that $\left(\beta,\mathrm{z1},\mathrm{z2}\right)\in]1,10]\times \llbracket0,29\rrbracket^{2}$ must be optimized simultaneously as part of the HP search as their effects on MSE are not independent.

\subsubsection*{C.3. Advantages over enforcing constraints after training}

\begin{figure}[b]
\begin{centering}
\includegraphics[width=0.5\textwidth]{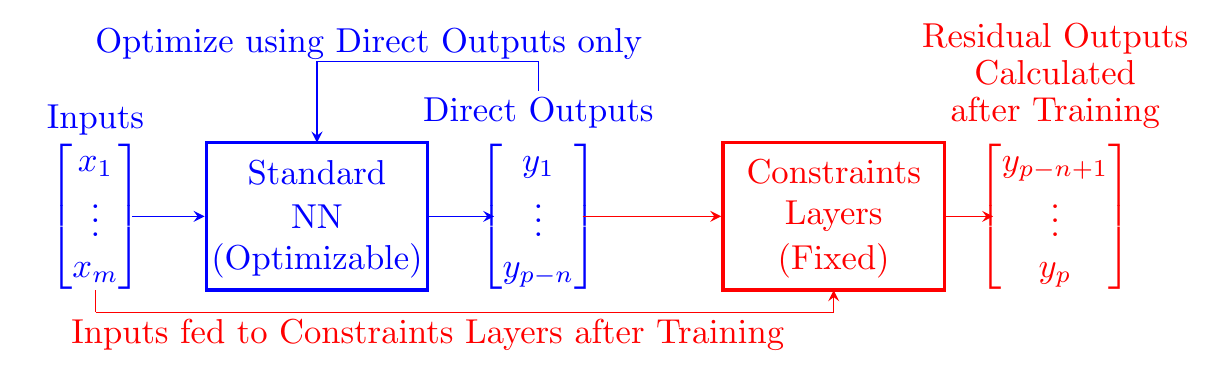}
\par\end{centering}
\caption{``Post-processing'' UCnet: A standard NN is trained on ``direct'' outputs only (blue) and the ``residual'' outputs are calculated after training, outside of the NN (red).}
\label{fig:C3_schematic}
\end{figure}

ACnets enforce constraints to within numerical precision \textit{during} training because they use constraints layers within their architecture. A more common method to ensure NNs satisfy constraints to within numerical precision is to enforce constraints \textit{after} training, referred to as ``post-processing'' \citep{Bolton2019} (abbreviated to pp hereafter): An unconstrained NN only predicts $\left(p-n\right)$ outputs and the $n$ other outputs are calculated as residuals \textit{after} the NN is trained (SM Figure \ref{fig:C3_schematic}). To clarify how ACnets compare to pp, we train 5 pp UCnets that have the same architecture as our 5 ACnets of lowest validation MSE from SM C.2.c, except for the constraints layers that are moved outside of the NN. By construction, pp UCnets enforce constraints to within numerical precision.

To evaluate the performance of pp UCnets, we first compare the MSE of their ``direct'' outputs with that of ACnets in SM Table \ref{tab:Postproc} over the validation (Line 1) and test (Line 5) datasets. Both ``direct'' MSEs are systematically to within 3$\% $ of each other, confirming that UCnets have been trained well enough to perform as well as ACnets. However, as ``residual'' outputs are calculated after training and hence not optimized using data, their MSEs are significantly larger for pp UCnets than for ACnets, except for the $q_{4}T_{4}{\beta}_{1.35}$ ACnet whose MSE for $\dot{T}_{4}\ $exceeds that of the corresponding pp UCnet (green spike in SM Figure \ref{fig:SE_wloss}j and \ref{fig:SE_wloss}l). Nevertheless, the $q_{8}T_{4}{\beta}_{1.49}$ ACnet examples proves that increasing $\beta $ can decrease the MSE for $\dot{T}_{4}\ $ so that it performs better than the corresponding pp UCnet (blue spike in SM Figure \ref{fig:SE_wloss}j and \ref{fig:SE_wloss}l). This affirms the superiority of ACnets over pp UCnets in predicting ``residual'' outputs as ACnets allow to decrease residual biases by weighting the loss function (SM B.3), which is unfeasible for pp UCnets whose loss function only includes ``direct'' outputs. That being said, the overall MSE of ACnets is always smaller but to within 3$\% $ of the overall MSE of pp UCnets. This means that the main advantage of ACnet over pp UCnets is exposing ``residual'' outputs to data during training, thereby improving corresponding predictions.   

\begin{table*}
\begin{centering}
{\small{}}%
\begin{tabular}{c|c|c|c|c|c|c}
{\small{}Dataset} & {\small{}MSE} & $q_{8}T_{4}\beta_{1.49}$ & $q_{3}T_{26}\beta_{1.15}$ & $q_{4}T_{4}\beta_{1.35}$ & $q_{5}T_{5}\beta_{1.72}$ & $q_{18}T_{28}\beta_{1.01}$\tabularnewline
\hline
\  & {\small{}$\mathrm{Direct}$} & {\small{}$1.6\ 10^{+02},\ 1.6\ 10^{+02}$} & {\small{}$1.6\ 10^{+02},\ 1.6\ 10^{+02}$} & {\small{}$1.6\ 10^{+02},\ 1.6\ 10^{+02}$} & {\small{}$1.6\ 10^{+02},\ 1.6\ 10^{+02}$} & {\small{}$1.6\ 10^{+02},\ 1.6\ 10^{+02}$}\tabularnewline
{\small{}Validation}  & {\small{}$\mathrm{z1}$} & {\small{}$4.4\ 10^{+01},\ 2.4\ 10^{+02}$} & {\small{}$5.9\ 10^{+01},\ 1.7\ 10^{+02}$} & {\small{}$8.5\ 10^{+01},\ 1.6\ 10^{+02}$} & {\small{}$6.6\ 10^{+01},\ 1.4\ 10^{+02}$} & {\small{}$8.2\ 10^{+02},\ 9.4\ 10^{+02}$}\tabularnewline
set  & {\small{}$\mathrm{z2}$} & {\small{}$4.6\ 10^{+01},\ 5.8\ 10^{+01}$} & {\small{}$1.6\ 10^{+02},\ 2.5\ 10^{+02}$} & {\small{}$8.4\ 10^{+01},\ 5.0\ 10^{+01}$} & {\small{}$9.3\ 10^{+01},\ 1.1\ 10^{+02}$} & {\small{}$1.5\ 10^{+02},\ 2.2\ 10^{+02}$}\tabularnewline
\  & {\small{}$\mathrm{Total}$} & {\small{}$1.6\ 10^{+02},\ 1.6\ 10^{+02}$} & {\small{}$1.6\ 10^{+02},\ 1.6\ 10^{+02}$} & {\small{}$1.6\ 10^{+02},\ 1.6\ 10^{+02}$} & {\small{}$1.6\ 10^{+02},\ 1.6\ 10^{+02}$} & {\small{}$1.6\ 10^{+02},\ 1.6\ 10^{+02}$}\tabularnewline
\cline{1-7}
\  & {\small{}$\mathrm{Direct}$} & {\small{}$1.6\ 10^{+02},\ 1.6\ 10^{+02}$} & {\small{}$1.6\ 10^{+02},\ 1.6\ 10^{+02}$} & {\small{}$1.6\ 10^{+02},\ 1.6\ 10^{+02}$} & {\small{}$1.6\ 10^{+02},\ 1.6\ 10^{+02}$} & {\small{}$1.5\ 10^{+02},\ 1.6\ 10^{+02}$}\tabularnewline
{\small{}Test}  & {\small{}$\mathrm{z1}$} & {\small{}$4.4\ 10^{+01},\ 2.4\ 10^{+02}$} & {\small{}$5.8\ 10^{+01},\ 1.7\ 10^{+02}$} & {\small{}$8.5\ 10^{+01},\ 1.6\ 10^{+02}$} & {\small{}$6.5\ 10^{+01},\ 1.4\ 10^{+02}$} & {\small{}$8.0\ 10^{+02},\ 9.1\ 10^{+02}$}\tabularnewline
set  & {\small{}$\mathrm{z2}$} & {\small{}$4.5\ 10^{+01},\ 5.7\ 10^{+01}$} & {\small{}$1.6\ 10^{+02},\ 2.5\ 10^{+02}$} & {\small{}$8.4\ 10^{+01},\ 5.0\ 10^{+01}$} & {\small{}$9.3\ 10^{+01},\ 1.1\ 10^{+02}$} & {\small{}$1.5\ 10^{+02},\ 2.2\ 10^{+02}$}\tabularnewline
\  & {\small{}$\mathrm{Total}$} & {\small{}$1.6\ 10^{+02},\ 1.6\ 10^{+02}$} & {\small{}$1.6\ 10^{+02},\ 1.6\ 10^{+02}$} & {\small{}$1.6\ 10^{+02},\ 1.6\ 10^{+02}$} & {\small{}$1.6\ 10^{+02},\ 1.6\ 10^{+02}$} & {\small{}$1.6\ 10^{+02},\ 1.6\ 10^{+02}$}\tabularnewline
\cline{1-7}
\hline
\end{tabular}
\par\end{centering}{\small \par}

\caption{Comparing ACnets to ``post-processed'' UCnets: MSE of ``direct'', ``residual'', and overall outputs. In each cell, the MSE for ACnets (left) is separated by a comma from the MSE for ``post-processed'' UCnets (right).\label{tab:Postproc}}
\end{table*}

\subsection*{D. Extending to Nonlinear Constraints: An Example}

In Section III of the main text, we focused on the mapping $\boldsymbol{x}\mapsto\boldsymbol{y}$,
where $\boldsymbol{x}\ $is given by Equation 10 and
$\boldsymbol{y}\ $is given by Equation 11. For this mapping,
the constraints -- conservation of column enthalpy, mass and radiation -- were conveniently \textit{linear} and given by $\boldsymbol{C}\begin{bmatrix}\boldsymbol{x} & \boldsymbol{y}\end{bmatrix}^{T}=\boldsymbol{0}$,
where the constraints matrix $\boldsymbol{C} $ is defined by Equation 12.

It is natural to wonder whether the approach demonstrated in this context is extendable to \textit{nonlinear} constraints as occur in many other physical systems such as enforcing kinetic energy conservation in a mapping from momentum to momentum time-tendencies.

Here, we show that indeed a similar strategy can be used to successfully
enforce \textit{nonlinear} physical constraints in a neural-network
emulating a different mapping $\boldsymbol{x_{0}}\mapsto\boldsymbol{y_{0}}$. For simplicity, we will work on the same
problem of thermodynamic convective parameterization for climate modeling. A nonlinearity is introduced in the formulation of the humidity variable. In SM D.1, we closely follow the steps of Figure 1 to enforce the resulting nonlinear constraints in a version of our architecture-constrained neural network that is adapted with ``conversion layers''. In SM D.2, we show
that the resulting network enforces nonlinear constraints to within excellent precision while maintaining high performance. Detailed documentation of the thermodynamics equations used for this network are given in SM D.3. The code for this network can be found at \url{https://github.com/tbeucler/CBRAIN-CAM/blob/master/notebooks/tbeucler_devlog/041_ACnet_Non_Linear.ipynb}.

\subsubsection*{D.1 Formulation\label{subsec:Formulation}}

Following Figure 1, the first step is to define the inputs and outputs of the mapping. Again, we map the local climate to how water vapor, liquid water, ice, and temperature are redistributed in the vertical,
but this time we describe water vapor in the input vector $\boldsymbol{x_{0}}\ $and
output vector $\boldsymbol{y_{0}}\ $using \textit{relative humidity} (definition below) instead of specific humidity. In addition to being a helpful illustration of how to handle nonlinearities, the physical context here is that there are reasons to think that reformulating moisture in this way will have advantages in improving the generalization of convection NNs trained in one climate to make predictions in another (e.g. warmer) one -- an active research frontier.  While specific humidity increases sharply
with temperature, relative humidity normalizes specific humidity so
 as to maintain values between $0\ $and $1\ $across climates, helpfully avoiding out-of-sample issues. 
 
 Mathematically,
relative humidity is defined as the ratio of the partial pressure
of water vapor $\boldsymbol{e}\boldsymbol{\left(p,q_{v}\right)}\ $to
its saturation value $\boldsymbol{e_{\mathrm{sat}}\left(T\right)}$,
and can be expressed analytically: 
\begin{equation}
\boldsymbol{\mathrm{RH}}\overset{\mathrm{def}}{=}\frac{\boldsymbol{e}\boldsymbol{\left(p,q_{v}\right)}}{\boldsymbol{e_{\mathrm{sat}}\left(T\right)}}\overset{\mathrm{def}}{=}\frac{R_{v}}{R_{d}}\frac{\boldsymbol{p}\boldsymbol{q_{v}}}{\boldsymbol{e_{\mathrm{sat}}\left(T\right)}},\label{eq:RH_def}
\end{equation}
where $R_{v}\approx461\mathrm{J\ kg^{-1}\ K^{-1}}\ $is the specific
gas constant for water vapor, $R_{d}\approx287\mathrm{J\ kg^{-1}\ K^{-1}}\ $is
the specific gas constant for dry air, $\boldsymbol{p}\ $(in units
Pa) is the total atmospheric pressure, $\boldsymbol{q_{v}}\ $(in
units kg/kg) is specific humidity, and $\boldsymbol{e_{\mathrm{sat}}\left(T\right)}\ $(in units
Pa) is
the saturation pressure of water vapor, whose analytic expression
in our case is given in SM D.3.
As $\boldsymbol{e_{\mathrm{sat}}}\boldsymbol{\left(T\right)}\ $increases
approximately exponentially with absolute temperature, relative humidity
is a strongly nonlinear function of specific humidity and temperature.
Therefore, the two first constraints in the present example, namely
conservation of mass and energy, are strongly nonlinear with respect
to the new relative humidity input and output. In summary, our input
vector $\boldsymbol{x_{0}}\ $is equal to $\boldsymbol{x}\ $(Equation
10) with $\boldsymbol{q_{v}}\ $replaced by $\boldsymbol{\mathrm{RH}}$,
our output vector $\boldsymbol{y_{0}}\ $is equal to $\boldsymbol{y}\ $(Equation
11) with $\boldsymbol{\dot{q}_{v}}\ $replaced by $\boldsymbol{\dot{\mathrm{RH}}}$,
and our constraints $\boldsymbol{c\left(x_{0},y_{0}\right)}=\boldsymbol{0}\ $are
strongly nonlinear:

\begin{equation} 
\boldsymbol{x_{0}}=\left[\begin{array}{ccc} \left(\boldsymbol{\mathrm{RH}},\boldsymbol{q_{l}},\boldsymbol{q_{i}},\boldsymbol{T},\boldsymbol{v},\boldsymbol{\mathrm{LS}},p_{s},S_{0}\right) & \mathrm{SHF} & \mathrm{LHF}\end{array}\right]^{T}, \label{eq:x0_def} 
\end{equation}

\begin{widetext}
\begin{equation} 
\boldsymbol{y_{0}}=\left[\begin{array}{ccccccccccccc} \boldsymbol{\dot{\mathrm{RH}}} & \boldsymbol{\dot{q}_{l}} & \boldsymbol{\dot{q}_{i}} & \boldsymbol{\dot{T}} & \boldsymbol{\dot{T}_{KE}} & \boldsymbol{\mathrm{lw}} & \boldsymbol{\mathrm{sw}} & \mathrm{LW_{t}} & \mathrm{LW_{s}} & \mathrm{SW_{t}} & \mathrm{SW_{s}} & P & P_{i}\end{array}\right]^{T}. \label{eq:y0_def} 
\end{equation}
\end{widetext}

\begin{figure}[b]
\begin{centering}
\includegraphics[width=0.5\textwidth]{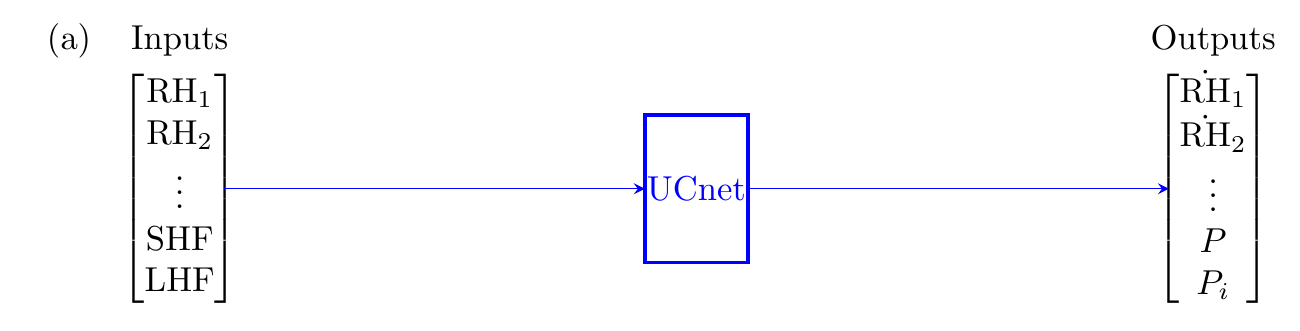}
\includegraphics[width=0.5\textwidth]{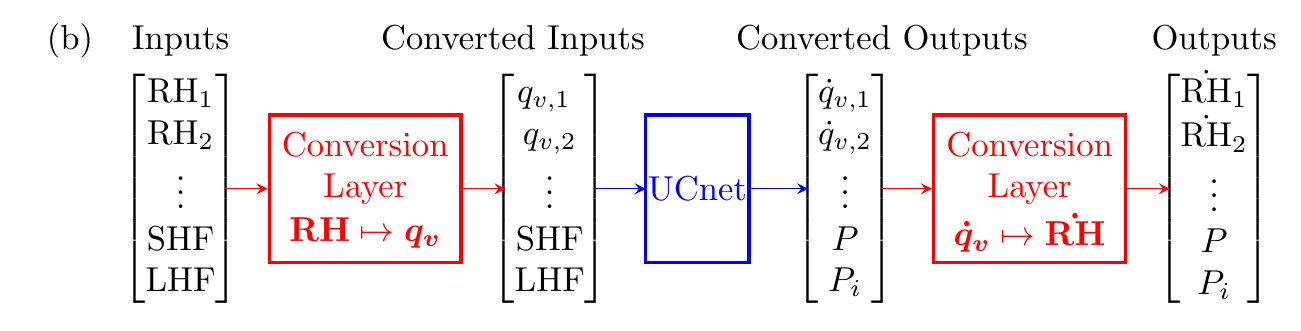}
\includegraphics[width=0.5\textwidth]{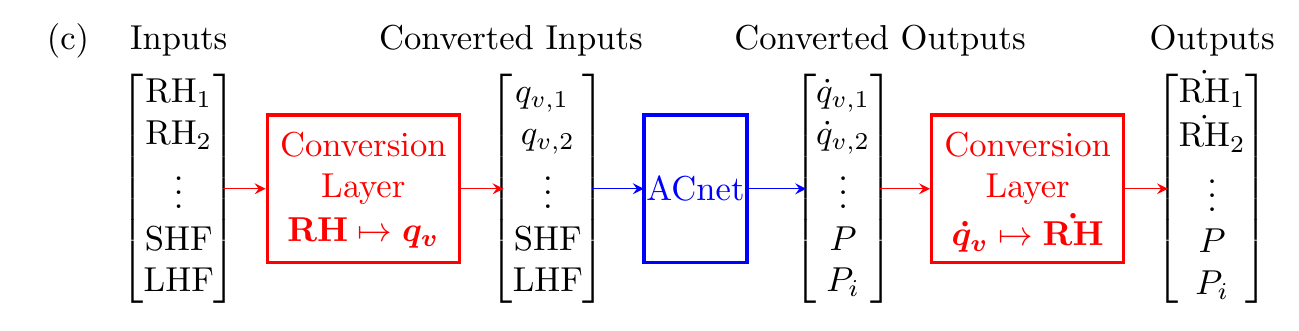}
\par\end{centering}
\caption{(a) $\mathrm{UCnet}$: Directly maps relative humidity inputs to relative humidity tendency outputs. \\
(b) $\mathrm{UCnet_{NL}}$: Inputs are converted from relative humidity to specific humidity before UCnet, after which outputs are converted back from specific humidity tendencies to relative humidity tendencies. \\
(c) $\mathrm{ACnet_{NL}}$: Same as (b) with ACnet instead of UCnet.}
\label{fig:D_schematic}
\end{figure}

The second step is to write $\boldsymbol{c}\ $as an explicit sum
of (1) $\boldsymbol{x}\ $only dependent on $\boldsymbol{x_{0}}$;
and (2) $\boldsymbol{y}\ $dependent on $\boldsymbol{x_{0}}\ $and
$\boldsymbol{y_{0}}$. In our case, we can choose the mapping $\boldsymbol{x}\mapsto\boldsymbol{y}\ $described
in the main paper, where conservation of mass and energy can be written
as an explicit sum of specific humidity tendencies. This natural choice yields Formulation 2 in Figure 1, i.e. a linearly-constrained mapping related to the nonlinearly constrained mapping of interest via a bijective function.
The last step is to build the network that maps $\boldsymbol{x_{0}}\ $to
$\boldsymbol{y_{0}}\ $while enforcing the constraints $\boldsymbol{c\left(x_{0},y_{0}\right)}=\boldsymbol{C}\begin{bmatrix}\boldsymbol{x} & \boldsymbol{y}\end{bmatrix}^{T}=\boldsymbol{0},$ which
are nonlinear with respect to $\boldsymbol{\left(x_{0},y_{0}\right)}\ $but
linear with respect to $\boldsymbol{\left(x,y\right)}$. For that
purpose, we augment the ACnet architecture described in Figure 2
with two ``conversion'' layers encoding the moist thermodynamics described in SM D.3:
\begin{enumerate}
\item A conversion layer $\left(\boldsymbol{\mathrm{RH}}\mapsto\boldsymbol{q_{v}}\right)\ $calculating
$\boldsymbol{q_{v}}\ $based on $\left(\boldsymbol{\mathrm{RH},T}\right)\ $to
convert $\boldsymbol{x_{0}}\ $to $\boldsymbol{x}\ $before ACnet.
\item A conversion layer $\left(\boldsymbol{\dot{q}_{v}}\mapsto\boldsymbol{\dot{\mathrm{RH}}}\right)\ $calculating
$\boldsymbol{\dot{\mathrm{RH}}}\ $based on $\left(\boldsymbol{\dot{q}_{v},\dot{T}}\right)\ $to
convert $\boldsymbol{y}\ $to $\boldsymbol{y_{0}}\ $after ACnet.
\end{enumerate}
The resulting network, which we refer to as the nonlinear ACnet ($\mathrm{ACnet_{NL}}\ $for
short), is depicted in SM Figure \ref{fig:D_schematic}c.  It is worth remarking that the idea of ``conversion layers'' should easily be adaptable to other physical systems with nonlinear constraints, such as ``converting'' velocity components into a kinetic energy upstream of an ACnet that enforces its linear conservation.

\subsubsection*{D.2 Results\label{subsec:Results}}

\begin{figure*}
\begin{centering}
\includegraphics[width=\textwidth]{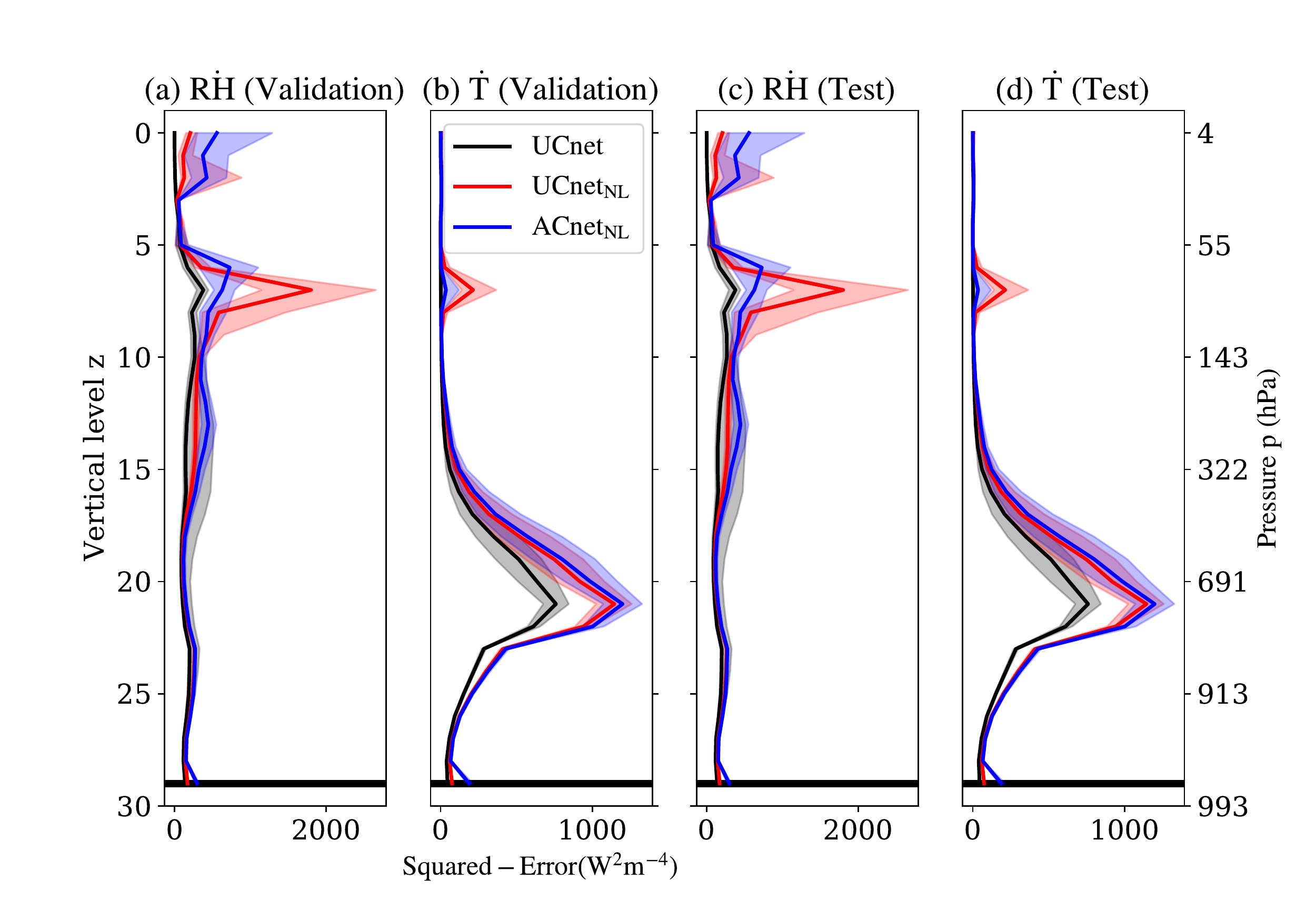}
\par\end{centering}
\caption{For $\mathrm{UCnet}$, $\mathrm{UCnet_{NL}}$, and $\mathrm{ACnet_{NL}}$: Squared error in convective moistening $\boldsymbol{\dot{q}_{v}} $ and heating $\boldsymbol{\dot{T}} $ versus pressure, for the validation and test sets. Each color represents a different NN type. We depict the median squared-error using full lines and shade the region between the first and third quartiles. Statistics use the full dataset and three NNs for each type (nine total). \label{fig:SE_NL}}
\end{figure*}

In this section, we compare the performance and constraints penalty of three types of NNs, all mapping $\boldsymbol{x_{0}}\mapsto\boldsymbol{y_{0}}$:
\begin{enumerate}
\item An unconstrained network (UCnet),
\item An unconstrained network using the two conversion layers $\left(\boldsymbol{\mathrm{RH}}\mapsto\boldsymbol{q_{v}}\right)\ $and
$\left(\boldsymbol{\dot{q}_{v}}\mapsto\boldsymbol{\dot{\mathrm{RH}}}\right)$,
referred to as the nonlinear UCnet ($\mathrm{UCnet_{\mathrm{NL}}}$
for short), to assess the effect of using conversion layers on optimization
independently of their actual purpose to manage nonlinear constraints,
\item A nonlinearly constrained network that exploits the two conversion layers to adapt the idea of ACnet to a nonlinear setting ($\mathrm{ACnet_{NL}}$).
\end{enumerate}
$\mathrm{UCnet_{NL}}$ and $\mathrm{ACnet_{NL}}$ are implemented using custom Tensorflow layers for the conversion layers. For each network type, we train three networks (total of 9 NNs) for 20 epochs using the RMSprop
optimizer and save the state of minimal validation loss to avoid overfitting. We report the performance and constraints penalty of $\mathrm{UCnet_{NL}}\ $and
$\mathrm{ACnet_{NL}}\ $for the validation and test sets in Table \ref{tab:D_results}. We complement Table \ref{tab:D_results} with the vertical profile
of the squared error in $\left(\boldsymbol{\dot{q}_{v}},\boldsymbol{\dot{T}}\right)\ $for
all three networks in SM Figure \ref{fig:SE_NL}.

The main result of this section is that we \textit{successfully enforced nonlinear constraints in NNs to excellent approximation}, as can be seen by the constraints penalty of $\mathrm{ACnet_{NL}} $ (Table VIII, bottom-right cell), which is 8 orders of magnitude smaller than that of $\mathrm{UCnet_{NL}} $ for both validation and test sets. Interestingly, the constraints penalty is more than one order of magnitude lower for $\mathrm{UCnet_{NL}} $ than for $\mathrm{UCnet} $, suggesting that unconstrained NNs are able without direction to approximate linear constraints better than strongly nonlinear constraints.

Table VIII also reveals how a reliance on conversion layers to handle nonlinear constraints does impact optimization with some training trade-offs in ways that are worth future research. As expected, UCnets are easiest to optimize. They exhibit mean-squared errors that are lower by a factor $\sim 1.5$ compared to NNs using ``conversion'' layers (Table VIII). This suggests that NNs with nonlinear conversion layers are somewhat harder to optimize. However, the absolute difference in MSE between $\mathrm{UCnet_{NL}} $ and $\mathrm{ACnet_{NL}} $ is smaller than $10 \textnormal{ W}^{2}\textnormal{m}^{-4} $, analogous to the small difference in MSE between UCnets and ACnets reported in Section III of the main text -- a price worth paying for the benefit of machine precision adherence to nonlinear constraints.

\begin{table*}
\begin{centering}
{\small{}}%
\begin{tabular}{c|c|c|c|c|c|c}
{\small{}Validation} & {\small{}Metric} & $\mathrm{UCnet}$ & $\mathrm{UCnet_{NL}}$ & $\mathrm{ACnet_{NL}}$\tabularnewline
\hline
{\small{}Validation} & {\small{}$\mathrm{MSE}$} & {\small{}$9.3\ 10^{+01}\pm4.4\ 10^{+02}$} & {\small{}$1.5\ 10^{+02}\pm5.9\ 10^{+02}$} & {\small{}$1.6\ 10^{+02}\pm5.7\ 10^{+02}$}\tabularnewline
set & {\small{}${\cal P}$} & {\small{}$7.2\ 10^{+04}\pm7.3\ 10^{+05}$} & {\small{}$3.9\ 10^{+03}\pm4.0\ 10^{+04}$} & {\small{}$2.1\ 10^{-04}\pm7.3\ 10^{-04}$}\tabularnewline
\cline{1-1}
{\small{}Test} & {\small{}$\mathrm{MSE}$} & {\small{}$9.3\ 10^{+01}\pm4.5\ 10^{+02}$} & {\small{}$1.5\ 10^{+02}\pm6.0\ 10^{+02}$} & {\small{}$1.5\ 10^{+02}\pm5.9\ 10^{+02}$}\tabularnewline
set & {\small{}${\cal P}$} & {\small{}$6.9\ 10^{+04}\pm7.3\ 10^{+05}$} & {\small{}$3.8\ 10^{+03}\pm4.2\ 10^{+04}$} & {\small{}$2.1\ 10^{-04}\pm7.1\ 10^{-04}$}\tabularnewline
\cline{1-1}
\hline
\end{tabular}
\par\end{centering}{\small \par}

\caption{NNs presented in SM Figure \ref{fig:D_schematic} (Ensemble mean of Mean MSE/Penalty $\pm $ Standard deviation). MSE is calculated using relative humidity tendencies, explaining its smaller values.}
\label{tab:D_results}
\end{table*}

\subsubsection*{D.3 Saturation pressure of water vapor\label{subsec:Saturation-pressure-of-water}}

In this section, we present the thermodynamics equations used to calculate
the saturation pressure of water vapor $\boldsymbol{e_{\mathrm{sat}}}\boldsymbol{\left(T\right)}$,
which allows to convert between specific and relative humidities.
The saturation pressure of water vapor can be found by integrating
the Clausius-Clapeyron equation with respect to temperature. Under
the microphysical assumptions of our fine-scale, cloud-resolving model
\citep{Khairoutdinov2003a}, it can be expressed analytically as:

\begin{widetext}
\begin{equation}
\boldsymbol{e_{\mathrm{sat}}\left(T\right)}=\begin{cases}
\boldsymbol{e_{\mathrm{liq}}\left(T\right)} & \boldsymbol{T}>T_{0}=273.16\mathrm{K}\\
\boldsymbol{e_{\mathrm{ice}}\left(T\right)} & \boldsymbol{T}<T_{00}=253.16\mathrm{K}\\
\boldsymbol{\omega e_{\mathrm{liq}}\left(T\right)}+\boldsymbol{\left(1-\omega\right)e_{\mathrm{ice}}\left(T\right)} & \boldsymbol{T}\in\left[T_{00},T_{0}\right]
\end{cases}.\label{eq:Sat_wat_vap_pressure}
\end{equation}
\end{widetext}

In Equation \ref{eq:Sat_wat_vap_pressure}, as temperature increases,
the saturation pressure of water vapor goes from the saturation vapor
pressure with respect to liquid $\boldsymbol{e_{\mathrm{liq}}}$,
given by the following polynomial approximation:
\begin{widetext}
\begin{equation}
\boldsymbol{e_{\mathrm{liq}}\left(T\right)}=100\mathrm{Pa}\times\sum_{i=0}^{8}a_{\mathrm{liq,i}}\left[\max\left(-193.15\mathrm{K},T-T_{0}\right)\right]^{i},
\end{equation}
\end{widetext}
where $\boldsymbol{a_{\mathrm{liq}}}\ $is a vector of length 9 containing
nonzero polynomial coefficients, to the saturation vapor pressure
with respect to ice $\boldsymbol{e_{\mathrm{ice}}}$, given by a different
polynomial approximation with temperature switches:
\begin{widetext}
\begin{equation}
\boldsymbol{e_{\mathrm{ice}}\left(T\right)}=\begin{cases}
\boldsymbol{e_{\mathrm{liq}}\left(T\right)} & \boldsymbol{T}>T_{0}\\
100\mathrm{Pa}\times\left\{ c_{\mathrm{ice},1}+\boldsymbol{{\cal C}\left(T\right)}\left[c_{\mathrm{ice},4}+c_{\mathrm{ice},5}\boldsymbol{{\cal C}\left(T\right)}\right]\right\}  & \boldsymbol{T}<T_{00}\\
100\mathrm{Pa}\times\sum_{i=0}^{8}a_{\mathrm{ice,i}}\left(\boldsymbol{T}-T_{0}\right)^{i} & \boldsymbol{T}\in\left[T_{00},T_{0}\right]
\end{cases},
\end{equation}  
\end{widetext}
where $\boldsymbol{{\cal C}\left(T\right)}\ $is a ramp function of
temperature given by:
\begin{equation}  
\boldsymbol{{\cal C}\left(T\right)}\overset{\mathrm{def}}{=}\max\left(c_{\mathrm{ice},2},\boldsymbol{T}-T_{0}\right),
\end{equation}  
and $\left(\boldsymbol{a_{\mathrm{ice}}},\boldsymbol{c_{\mathrm{ice}}}\right)\ $are
vectors of length 9 and 5 containing nonzero elements, respectively.
Between temperatures of $T_{00}\ $and $T_{0}$, the saturation pressure
of water vapor is a weighted mean of $\boldsymbol{e_{\mathrm{liq}}}\ $and
$\boldsymbol{e_{\mathrm{ice}}}$, where the weight $\omega\ $is a
linear function of the absolute temperature:
\begin{equation}  
\boldsymbol{\omega}\overset{\mathrm{def}}{=}\frac{\boldsymbol{T}-T_{00}}{T_{0}-T_{00}}.
\end{equation}  
The reader interested in the numerical details of this calculation
is referred to our implementation of relative humidity at \url{https://github.com/tbeucler/CBRAIN-CAM/blob/master/notebooks/tbeucler_devlog/041_ACnet_Non_Linear.ipynb}.

\subsection*{E. Enforcing Inequality Constraints in Architecture-Constrained Networks}

In this section, we briefly discuss how to enforce inequality constraints in ACnets using a concrete example: the positivity of liquid water concentration in this letter's NN parameterization of convection. This inequality requires the liquid water concentration $q_{l,z}\left(t\right) $ at a given vertical level $z $ and at the current timestep $t $ to be positive. In practice, the liquid water concentration $q_{l,z}\left(t\right) $ is obtained from the liquid water concentration $q_{l,z}\left(t-1\right) $ and the liquid water tendency $\dot{q}_{l,z}\left(t-1\right) $ at the previous timestep $t-1 $ through time-stepping, which means that we can write the positivity constraint as:
\begin{equation}
q_{l,z}\left(t\right)=\overbrace{q_{l,z}\left(t-1\right)}^{\textnormal{Input}}+\Delta t\times\overbrace{\dot{q}_{l,z}\left(t-1\right)}^{\textnormal{Output}}\geq0,
\end{equation}
where we note that $q_{l,z}\left(t-1\right) $ is an NN input while $\dot{q}_{l,z}\left(t-1\right) $ is an NN output.

To enforce this inequality constraint in ACnet, we choose $\dot{q}_{l,z}\left(t-1\right) $ to be a ``direct'' NN output and add an inequality constraints layer (ICL) before the constraints layer (CL). Although any positive-definite activation function can be used, a possible implementation of (ICL) uses the rectified linear unit (ReLU) activation function:
\begin{widetext}
\begin{equation}
\underbrace{\dot{q}_{l,z}\left(t-1\right)}_{\textnormal{After\ \ensuremath{\left(\textnormal{ICL}\right)}}}=\mathrm{ReLU}\left[\underbrace{\dot{q}_{l,z}\left(t-1\right)}_{\textnormal{Before\ \ensuremath{\left(\textnormal{ICL}\right)}}}+\Delta t^{-1}q_{l,z}\left(t-1\right)\right]-\Delta t^{-1}q_{l,z}\left(t-1\right),
\end{equation}
\end{widetext}
so that the liquid water tendency after (ICL) yields a positive liquid water concentration $q_{l,z}\left(t\right) $ at the current timestep. Since the inequality constraints layer (ICL) comes before the constraints layer, which do not change $q_{l,z}\left(t\right) $, we can still enforce the equality constraints $\left(\cal{C}\right) $ to within machine precision. Finally, note that the previous example can be generalized to nonlinear analytic constraints involving less than $\left(p-n\right) $ inputs, allowing ACnets to enforce a broad range of inequality constraints.

\bibliography{supp.bbl}